\documentclass[12pt, draftclsnofoot, onecolumn]{IEEEtran}

\IEEEoverridecommandlockouts

\usepackage{cite}
\usepackage{graphicx}
\usepackage[cmex10]{amsmath}
\allowdisplaybreaks
\usepackage{amssymb}
\usepackage{array}
\usepackage{psfrag}
\usepackage{arydshln}
\usepackage{pifont}
\usepackage{cancel}
\usepackage{wasysym}
\usepackage{float}
\usepackage{amsthm}
\theoremstyle{plain}
\usepackage{bm}

\usepackage[mathcal]{euscript}

\usepackage{color}

\usepackage{multirow}
\ifx\pdfoutput\undefined
\usepackage[hypertex]{hyperref}
\else
\usepackage[pdftex,hypertexnames=false]{hyperref}
\fi

\usepackage{footnote}
\makesavenoteenv{tabular}

\newcommand{\hFoo}{{\kern.15em{}_{1}\kern-.05em {\sf F}_{1}}}

\newcommand{\hFooa}{{\kern.15em{}_{1}\kern-.05em {\sf F}_{1}^{(1)}}}

\newcommand{\hFoob}{{\kern.15em{}_{1}\kern-.05em {\sf F}_{1}^{(2)}}}

\newtheorem{lemma}{Lemma}
\newtheorem{theorem}{Theorem}

\newtheorem{remark}{Remark}

\newcommand{\mb}{\mathbf}

\newcommand{\uh}{{\mb{h}}}
\newcommand{\uha}{{\uh_1}}
\newcommand{\ufa}{{\uf_1}}
\newcommand{\ufaH}{\uf_1^{\sf H}}

\newcommand{\uhaH}{\uh_1^{\sf H}}

\newcommand{\uhd}{{\mb{h}_{\text{d}}}}
\newcommand{\uhda}{{\mb{h}_{\text{d},1}}}

\newcommand{\ufatilde}{\widetilde{\uf}_1}
\newcommand{\ufatildeH}{\widetilde{\uf}_1^{\sf H}}

\newcommand{\ufda}{{\mb{f}_{\text{d},1}}}

\newcommand{\uhra}{{\mb{h}_{\text{r},1}}}
\newcommand{\ufra}{{\mb{f}_{\text{r},1}}}

\newcommand{\umu}{\pmb{\mu}}

\newcommand{\uPhi}{{\pmb{\Phi}}}

\newcommand{\uTheta}{{\pmb{\Theta}}}

\newcommand{\uPsi}{{\pmb{\Psi}}}

\newcommand{\unu}{{\pmb{\nu}}}

\newcommand{\ur}{{\mb{r}}}

\newcommand{\up}{{\mb{p}}}
\newcommand{\urtilde}{\widetilde{\ur}}

\newcommand{\um}{{\mb{m}}}

\newcommand{\ua}{{\mb{a}}}
\newcommand{\uaH}{{\mb{a}}^{\sf H}}

\newcommand{\uA}{{\mb{A}}}

\newcommand{\uAH}{{\mb{A}}^{\sf H}}
\newcommand{\uAinvH}{{{\mb{A}}^{-{\sf H}}}}
\newcommand{\uQ}{{\mb{Q}}}

\newcommand{\ub}{{\mb{b}}}

\newcommand{\ubtilde}{\widetilde{\mb{b}}}

\newcommand{\uc}{{\mb{c}}}

\newcommand{\uf}{{\mb{f}}}

\newcommand{\uH}{{\mb{H}}}
\newcommand{\uHH}{{\mb{H}}^{\sf H}}

\newcommand{\uHb}{{\mb{H}_2}}
\newcommand{\uHbH}{{\mb{H}_2^{\sf H}}}

\newcommand{\uGb}{{\mb{G}_2}}
\newcommand{\uGbH}{{\mb{G}_2^{\sf H}}}

\newcommand{\uFb}{{\mb{F}_2}}
\newcommand{\uFbH}{{\mb{F}_2^{\sf H}}}

\newcommand{\uFdb}{{\mb{F}_{\text{d},2}}}

\newcommand{\uFrb}{{\mb{F}_{\text{r},2}}}

\newcommand{\uErb}{{\mb{E}_{\text{r},2}}}
\newcommand{\uErbH}{{\mb{E}_{\text{r},2}^{\sf H}}}

\newcommand{\uEb}{{\mb{E}_2}}
\newcommand{\uEbH}{{\mb{E}_2^{\sf H}}}

\newcommand{\uHw}{{{\mb{H}}}_{\text{w}}}

\newcommand{\uEwb}{{{\mb{E}}}_{\text{w},2}}

\newcommand{\uG}{{\mb{G}}}

\newcommand{\uFd}{{\uF_{\text{d}}}}

\newcommand{\uHd}{{\uH_{\text{d}}}}

\newcommand{\uHdb}{{\uH_{\text{d},2}}}

\newcommand{\uGdb}{{\uG_{\text{d},2}}}

\newcommand{\uEdb}{{\uE_{\text{d},2}}}

\newcommand{\uGbtilde}{{\widetilde{\uG}_{2}}}

\newcommand{\uHdH}{{\uH_{\text{d}}^{\sf H}}}

\newcommand{\uHdn}{{\uH_{\text{d,n}}}}

\newcommand{\uHr}{{\uH_{\text{r}}}}

\newcommand{\uHrH}{{\uH_{\text{r}}^{\sf H}}}

\newcommand{\uHrb}{{\uH_{\text{r},2}}}

\newcommand{\uHrn}{{\uH_{\text{r,n}}}}
\newcommand{\uHrnH}{{\uH_{\text{r,n}}^{\sf H}}}

\newcommand{\NT}{{N_{\text{T}}}}
\newcommand{\NTsq}{{N_{\text{T}}^2}}
\newcommand{\NTcb}{{N_{\text{T}}^3}}
\newcommand{\NR}{{N_{\text{R}}}}

\newcommand{\NI}{{N_{\text{I}}}}

\newcommand{\uP}{{\mb{P}}}
\newcommand{\uR}{{\mb{R}}}

\newcommand{\uRT}{{\mathbf{R}_{\text{T}}}}

\newcommand{\uRTK}{{\mathbf{R}_{\text{T,}K}}}

\newcommand{\uRTKsqrt}{{\mathbf{R}^{1/2}_{\text{T,}K}}}

\newcommand{\uy}{{\mb{y}}}

\newcommand{\un}{{\mb{n}}}

\newcommand{\mzero}{\mb{0}}

\newcommand{\mbR}{{\mb{R}}}

\newcommand{\uE}{{\mb{E}}}
\newcommand{\uF}{{\mb{F}}}

\newcommand{\uU}{{\mb{U}}}

\newcommand{\uUb}{{\mb{U}}_2}

\newcommand{\uUbtilde}{\widetilde{{\mb{U}}}_2}

\newcommand{\uV}{{\mb{V}}}

\newcommand{\uVtilde}{\widetilde{\mb{V}}}
\newcommand{\uVtildeH}{\widetilde{\mb{V}}^{\sf H}}

\newcommand{\uT}{{\mb{T}}}

\newcommand{\uXi}{{\pmb{\Xi}}}
\newcommand{\mbI}{\mb{I}}

\newcommand{\RTinvaa}{\left[\mathbf{R}_{\text{T},K}^{-1}\right]_{1,1}}

\newcommand{\RTaa}{\mathbf{R}_{{\text{T},K}_{11}}}
\newcommand{\rTaa}{r_{{\text{T},K}_{11}}}

\newcommand{\RTbb}{\mathbf{R}_{{\text{T},K}_{22}}}

\newcommand{\RTbbinva}{\mathbf{R}_{{\text{T},K}_{22}}^{-1}}

\newcommand{\RTab}{\mathbf{R}_{{\text{T},K}_{12}}}

\newcommand{\RTba}{\mathbf{R}_{{\text{T},K}_{21}}}
\newcommand{\rTba}{\mathbf{r}_{{\text{T},K}_{21}}}
\newcommand{\rTbaH}{\mathbf{r}_{{\text{T},K}_{21}}^{\sf H}}

\newcommand{\uqsimple}{\mathbf{q}}

\newcommand{\cmark}{\ding{51}}
\newcommand{\xmark}{\ding{55}}

\def\hF{{\kern.15em{}_{1}\kern-.05emF_{1}}}

\title{Exact {\sf ZF} Analysis and Computer-Algebra-\\Aided Evaluation in Rank-$1$ LoS Rician Fading}

\author{
Constantin Siriteanu\thanks{C.~Siriteanu is with the Graduate School of Information Science and Technology, Osaka University, Osaka 565-0871, Japan.}, 
Akimichi Takemura\thanks{A.~Takemura was with the Graduate School of Information Science and Technology, University of Tokyo, Tokyo 113-8656, Japan. He is now with the Center for Data Science Education and Research, Shiga University, Hikone, Japan.},
Christoph Koutschan\thanks{C. Koutschan is with the Johann Radon Institute for Comput. and Applied Math., Austrian Academy of Sciences, Linz.},\\
Satoshi Kuriki\thanks{S.~Kuriki is with the Institute of Statistical Mathematics, Tachikawa, Tokyo, Japan.}, 
Donald St. P. Richards\thanks{D. St.~P.~Richards is with the Department of Statistics, Pennsylvania State University, University Park, PA, USA.},
Hyundong Shin\thanks{H.~Shin is with the Department of Electronics and Radio Engineering, Kyung Hee University, Yongin, Republic of Korea.}}

\begin{document}

\maketitle

\begin{abstract}
We study zero-forcing detection ({\sf ZF}) for multiple input/multiple output (MIMO) spatial multiplexing under transmit-correlated Rician fading for an $ \NR \times \NT $ channel matrix with rank-$1$ line-of-sight (LoS) component.
By using matrix transformations and multivariate statistics, our exact analysis yields the signal-to-noise ratio moment generating function (m.g.f.) as an infinite series of gamma distribution m.g.f.'s and analogous series for {\sf ZF} performance measures, e.g., outage probability and ergodic capacity. 
However, their numerical convergence is inherently problematic with increasing Rician $ K $-factor, $ \NR $, and $ \NT $.
We circumvent this limitation as follows.
First, we derive differential equations satisfied by the performance measures with a novel automated approach employing a computer-algebra tool which implements {{Gr{\"o}bner basis}} computation and creative telescoping.
These differential equations are then solved with the holonomic gradient method (HGM) from initial conditions computed with the infinite series. 
We demonstrate that HGM yields more reliable performance evaluation than by infinite series alone and more expeditious than by simulation, for realistic values of $ K $, and even for $ \NR $ and $ \NT $ relevant to large MIMO systems.
We envision extending the proposed approaches for exact analysis and reliable evaluation to more general Rician fading and other transceiver methods.
\end{abstract}




\section{Introduction}

\label{section_introduction}

\subsection{Background, Motivation, and Scope}

The performance of multiple input/multiple output (MIMO) wireless communications systems has remained under research focus as the multiantenna architectures that attempt to harvest MIMO gains have continued to evolve, e.g., from single-user MIMO, to multi-user and distributed MIMO, and, most recently, to massive or large MIMO\cite{tse_book_05},\cite{gesbert_sp_05},\cite{ngo_tvt_13},\cite{rusek_spm_13},\cite{hoydis_jsac_13},\cite{lu_stsp_14},\cite{bjornsson_cm_15}.


As the numbers of transmitting and receiving antennas, herein denoted with $ \NT $ and $ \NR $, respectively, have increased in seeking higher array, diversity, and multiplexing gains\cite[pp.~72,~64,~385]{tse_book_05}, transceiver processing complexity has also increased. 
For spatial multiplexing transmission, linear detection methods\cite{ngo_tvt_13},\cite{hoydis_jsac_13},\cite{lu_stsp_14}, e.g., zero-forcing detection ({\sf ZF}) and minimum mean-squared-error detection ({\sf MMSE}), are attractive because of their relatively-low complexity order $ {\cal{O}} (\NR \NT + \NR \NTsq + \NTcb) $\cite{lu_stsp_14} and their good performance for $ \NR \gg \NT $, as the columns of the $ \NR \times \NT $ channel matrix $ \uH $ tend to become independent\cite{lu_stsp_14}.  


For increased practical relevance, MIMO channel model complexity has also been growing, and, with it, the difficulties of MIMO performance analysis and numerical evaluation.
Thus, early {\sf ZF} research assumed zero-mean, i.e., Rayleigh fading, for the elements of $ \uH $, which enabled relatively simple analysis and evaluation\cite{winters_tcom_94},\cite{gore_cl_02},\cite{kiessling_spawc_03}. 
Recently, various cases of nonzero-mean $ \uH $, i.e., Rician fading, have rendered increasingly more difficult the analysis and evaluation for several transceiver methods\cite{nabar_twc_05},\cite{kang_twc_06},\cite{jin_tit_07},\cite{mckay_tcomm_09},\cite{matthaiou_tsp_11},\cite{li_spawc_14energy},\cite{zhang_jstsp_14},\cite{siriteanu_tvt_11},\cite{siriteanu_twc_13},\cite{siriteanu_twc_14},\cite{siriteanu_twc_SC_14}.

Rician fading can occur due to line-of-sight (LoS) propagation, in  indoor, urban, and suburban scenarios, as shown by the WINNER II channel measurements\cite[Section~2.3]{winner_d_1_1_2_v_1_2}.
WINNER II\cite[Table~5.5]{winner_d_1_1_2_v_1_2}  has also characterized as lognormal the distributions of 1) the Rician $ K $-factor, which determines the strength of the channel mean vs.~standard deviation\cite[p.~37]{tse_book_05}, and 2) the azimuth spread (AS), which determines the antenna correlation\cite[p.~136]{siriteanu_thesis_06}. 
An ability to evaluate MIMO performance over the range of realistic values of $ K $ and AS is useful, e.g., in averaging over their distributions, which has only rarely been attempted before\cite{siriteanu_tvt_11}.

Consequently, we focus herein on evaluating MIMO {\sf ZF} under transmit-correlated Rician fading. 
For tractable analysis we assume as in\cite{matthaiou_tsp_11},\cite{li_spawc_14energy} that the LoS or deterministic component of $ \uH $ satisfies $ \text{rank} (\uHd) = r = 1 $. 
Whereas for LoS propagation $ r $ can take any value from $ 1 $ to $ \NT $\cite{torkildson_twc_11},\cite{brady_tap_13},\cite{heath_jstsp_16},\cite{sayeed_tsp_02}, small antenna apertures, relatively-low carrier frequency, or large transmitter-receiver distance, as in conventional point-to-point deployments\cite{tse_book_05},\cite{li_spawc_14energy}, are likely to yield $ \uHd $ as outer product of array response vectors\cite[Eq.~(7.29), p. 299]{tse_book_05}, i.e., $ r = 1 $.

Our future work shall consider Rician fading with $ r > 1 $ for {\sf ZF} and MMSE. 
Higher $ r $, which improves $ \uH $ conditioning, i.e., MIMO performance, is becoming increasingly more relevant due to envisioned LoS millimeter-wave applications\cite{torkildson_twc_11}.
MMSE is appealing because it outperforms ZF. 
Also, we shall tackle more general statistical fading models that can characterize more modern MIMO deployment types\cite{zhang_jstsp_14}.
Finally, for millimeter waves and massive MIMO, we shall pursue beamspace channel matrix representation and signal processing\cite{sayeed_tsp_02},\cite{brady_tap_13},\cite{sayeed_globecom_13}.

\subsection{Limitations of Relevant Previous Work on MIMO {\sf ZF}}
\label{section_limitations_previous_work}

Historically, the study of MIMO {\sf ZF} commenced with that for uncorrelated Rayleigh fading from\cite{winters_tcom_94}.
The case of transmit-correlated Rayleigh fading was elucidated in\cite{gore_cl_02},\cite{kiessling_spawc_03}. 
For Rician fading, previous studies assumed certain values for $ r $ and/or proceeded by approximation:
\begin{itemize}
\item Rician fading only for 1) the intended stream, i.e., \textit{Rician--Rayleigh fading}, which is a special case with $ r = 1 $, or 2) the interfering streams, i.e., \textit{Rayleigh--Rician fading}, whereby $ r = \NT - 1$; these cases may arise in heterogeneous networks. Then, we derived in\cite{siriteanu_twc_13} exact infinite-series expressions for performance measures, e.g., the average error probability, outage probability, and ergodic capacity (i.e., rate\cite{matthaiou_tsp_11},\cite{zhang_jstsp_14}) --- more details below.
\item Rician fading for all streams, i.e., \textit{full-Rician} fading, for the special case with $ r = 1 $. 
Early works --- see\cite{siriteanu_tvt_11},\cite{matthaiou_tcomm_10} and references therein --- used an approximation of the cumbersome noncentral-Wishart distribution of $ \uHH \uH $ with a central-Wishart distribution of equal mean, which has yielded a simple gamma distribution for the signal-to-noise ratio ({\sf SNR}). 
Then,\cite{siriteanu_tvt_11},\cite{siriteanu_twc_SC_14} reveal that $ r = 1 $ does not ensure consistent approximation accuracy\footnote{Only very careful usage in\cite{siriteanu_tvt_11} helped average the performance over WINNER II distributions of $ K $ and AS for $ r = 1 $.} and $ r>1 $ can render it useless.
Recently, bounding techniques have yielded --- only for uncorrelated fading --- the simple sum rate bounds in\cite[Eqs.~(55)--(58)]{matthaiou_tsp_11} that become accurate at high SNR.
\item Rician fading, $ \forall r = 1, \, \cdots, \NT $. For this most general case, exact sum-rate expressions for $ \NR \rightarrow \infty $ and approximations for finite $ \NR $ were derived in\cite{zhang_jstsp_14}.
\end{itemize}

For Rician--Rayleigh fading, we have recently analyzed and evaluated {\sf ZF} exactly in\cite{siriteanu_twc_13},\cite{siriteanu_twc_14}. 
In\cite{siriteanu_twc_13}, we expressed the {\sf SNR}  moment generating function (m.g.f.) in terms of the confluent hypergeometric function $ {\hFoo(\cdot, \cdot, \sigma)} $\cite[Eq.~(31)]{siriteanu_twc_13}, where $ \sigma  \propto K \NR \NT $.
Thereafter, its well-known expansion around $ \sigma_0 = 0 $\cite[Eq.~(30)]{siriteanu_twc_13} yielded an infinite series of gamma distribution m.g.f.'s\cite[Eq.~(37)]{siriteanu_twc_13}.
Finally, inverse-Laplace transformation and integration yielded analogous series for the {\sf SNR}  probability density function (p.d.f.), average error probability, outage probability, and ergodic capacity\cite[Eqs.~(39),~(58),~(69),~(71)]{siriteanu_twc_13}.
However, beside complicating the analysis, the Wishart distribution noncentrality induced by Rician fading also leads to numerical divergence for these series  with increasing $ K $, $\NR$, and $ \NT $\cite[Section~V.F]{siriteanu_twc_13}.
In\cite{siriteanu_twc_14}, we overcame this limitation by using the fact that $ {\hFoo(\cdot, \cdot, \sigma)} $ is a \textit{holonomic function}\footnote{Other examples: rational functions, logarithm, exponential, sine, special functions (orthogonal polynomials,
Bessel\cite[p.~41]{koutschan_thesis_09}).}, i.e., it satisfies a differential equation\cite[Eq.~(27)]{siriteanu_twc_14} with polynomial coefficients with respect to (w.r.t.) $ \sigma $. 
Starting from this differential equation, a difficult by-hand derivation produced differential equations for the {\sf SNR}  m.g.f.~and then for the  {\sf SNR} p.d.f., via inverse-Laplace transform.
Thereafter, we computed reliably the p.d.f.~at realistic values of $ K $ --- but only for relatively small $ \NR $ and $ \NT $ --- by numerically solving its differential equations from initial conditions computed with the infinite series for small $ K $. 
This approach is known as the \textit{holonomic gradient method} (HGM) because, at each step, the function value is updated with the differential gradient\cite[Sec.~IV.B]{siriteanu_twc_14}. 
Finally, in\cite{siriteanu_twc_14}, the {\sf SNR}  p.d.f.~computed with HGM was numerically integrated to evaluate performance measures, i.e., the outage probability and ergodic capacity.

Thus, on the one hand, our exact studies for $ r=1 $ in\cite{siriteanu_twc_13},\cite{siriteanu_twc_14} are limited by the following:
\begin{itemize}
\item Nonfull-Rician (i.e., only Rician--Rayleigh) fading assumption.
\item Tedious by-hand derivations of the {\sf SNR}  m.g.f.~and p.d.f.~differential equations.
\item Time-consuming numerical integration of the p.d.f.~for performance measure evaluation.
\item HGM not tried for large $ \NR $ and $ \NT $, e.g., as relevant for large MIMO systems\cite{hoydis_jsac_13},\cite{bjornsson_cm_15}.
\end{itemize}
On the other hand, only approximations exist for full-Rician fading and $ r = 1 $\cite{matthaiou_tsp_11},\cite{zhang_jstsp_14},\cite{siriteanu_tvt_11}.


\subsection{Problem Tackled in the Current Work; Exact Analysis and Evaluation Approaches}
\label{section_problem}

To the best of our knowledge, the performance of MIMO {\sf ZF} has not yet been studied \textit{exactly} under full-Rician fading even for $ r = 1 $.
We pursue this study herein, as follows.

First, upon applying a sequence of matrix transformations and results from multivariate statistics, we obtain several theoretical results that help express exactly the {\sf SNR} m.g.f.~as an infinite series with terms in $ {\hFoo(\cdot, \cdot, \cdot)} $.
Thus, the m.g.f.~can be rewritten as a double-infinite series of gamma distribution m.g.f.'s, which readily yields analogous series for the {\sf SNR}  p.d.f.~and for the performance measures.
Then, they are recast as a generic single-infinite series. However, its truncation is found to incur numerical divergence with increasing $ K $, $\NR$, and $ \NT $. 
Consequently, as in\cite{siriteanu_twc_13}, it is necessary to derive satisfied differential equations and apply HGM.

Because by-hand derivation of differential equations for our generic series appears intractable, we resort to a novel automated derivation approach using the \texttt{HolonomicFunctions}   package written earlier by one of the authors\cite{koutschan_thesis_09},\cite{koutschan_tr_10b} and implementing recent advances in computer algebra. 
It exploits, for holonomic functions, closure properties\cite[Section~IV.C]{siriteanu_twc_14},\cite{koutschan_thesis_09}, the algebraic concept of \textit{{Gr{\"o}bner bases}}\footnote{\label{footnote_Grobner}Buchberger's algorithm\cite{buchberger_thesis_65} for {Gr{\"o}bner basis} computation
specializes, for example, to the Euclidean algorithm when applied to univariate polynomials,
and to Gaussian elimination when applied to linear polynomials in several
variables\cite{lin_tcs_08}. {Gr{\"o}bner bases} have helped solve communications optimization problems cast as systems of polynomial equations, e.g., for interference alignment\cite{geng_jsac_14}, coding gain maximization in space--time coding\cite{fozunbal_tit_04}; other relevant applications are listed in\cite{lin_tcs_08}.},\cite{lin_tcs_08}, and \textit{creative telescoping} algorithms\cite[Ch.~3]{koutschan_thesis_09} to systematically deduce differential equations for their addition, multiplication, composition, and integration.
This computer-algebra-aided approach readily yields differential equations not only for the {\sf SNR}  m.g.f.~and p.d.f., but also for the outage probability and ergodic capacity.


Finally, we evaluate {\sf ZF} performance measures by HGM, i.e., by solving the obtained differential equations starting from initial conditions computed with the infinite series.

\subsection{Contributions}
\label{section_contributions}

Compared to previous MIMO {\sf ZF} work by us and others, herein we: 
\begin{itemize}
\item Tackle full-Rician fading with $ r = 1 $ in a new exact analysis that reveals that the {\sf SNR}  distribution is an infinite mixture of gamma distributions. This {\sf SNR}  distribution yields insight into the effect of channel matrix statistics (mean, correlation) on performance, and helps reassess the approximation with the gamma distribution we studied in\cite{siriteanu_tvt_11},\cite{siriteanu_twc_SC_14}.
\item Use computer algebra to automate deductions of differential equations also for performance measures and, thus, also avoid time-consuming numerical integration of the {\sf SNR}  p.d.f.. 
\item Demonstrate that HGM yields accurate performance evaluation for realistic values for $ K $, and even for large $ \NR $ and $ \NT $, unlike the infinite series alone and faster than by simulation. 
\item Exactly average the {\sf ZF} performance over WINNER II distributions of $ K $ and AS.
\end{itemize}


\subsection{Paper Organization}
Section~\ref{section_system_channel_model} introduces our model.
Section~\ref{section_MGF_unconditioned_SNR} employs matrix transformations and multivariate statistics to express exactly the m.g.f.~of the {\sf ZF} {\sf SNR}.
Section~\ref{section_infinite_series_for_all} derives a generic infinite series for the {\sf SNR}  m.g.f.~and p.d.f., as well as for {\sf ZF} performance measures.
Section~\ref{section_differential_equations_all} describes the automated derivation of differential equations, which has been accomplished with \texttt{HolonomicFunctions} commands as shown in\cite{koutchan_Mathnb_LOS}. 
Finally, Section~\ref{section_numerical_results} presents numerical results obtained by simulation, series truncation, and HGM.
The Appendix shows some proofs and derivation details.



\subsection{Notation}
\label{section_notation}
\begin{itemize}
\item Scalars, vectors, and matrices are represented with lowercase italics, lowercase boldface, and uppercase boldface, respectively, e.g., $ y $, $ \uh $, and $ \uH $; the statement $ \uH \doteq \NR \times \NT $ indicates $ \NR $ rows and $ \NT $ columns for $ \uH $; zero vectors and matrices of appropriate dimensions are denoted with $ \mzero $; superscripts $ \cdot^{\sf T} $ and $ \cdot^{\sf H} $ stand for transpose and Hermitian (i.e., complex-conjugate) transpose; $ \mbI_N $ is the $ N \times N $ identity matrix.
\item $ [\cdot]_{i} $ is the $i$th element of a vector; $ [\cdot]_{i,j} $, $ [\cdot]_{i,\bullet} $, and $ [\cdot]_{\bullet,j} $ indicate the $i,j$th element, $i$th row, and $j$th column of a matrix; $ \| \uH \|^2 = \sum_{i=1}^{\NR} \sum_{j=1}^{\NT} | [\uH]_{i,j} |^2 $ is the squared Frobenius norm.
\item $ i = 1 : N $ stands for the enumeration $ i = 1, \, 2, \, \ldots, \, N $; $ \otimes $ stands for the Kronecker product\cite[p.~72]{gentle_book_07} ; $ \propto $ stands for `proportional to'; $ \Rightarrow $ stands for logical implication.
\item $ \uh \sim {\cal{CN}}_{\NR} \left( \uhd, \uR \right) $ denotes an $ \NR \times 1 $ complex-valued circularly-symmetric Gaussian vector with mean $ \uhd $ and covariance matrix $ \uR $; an $ \NR \times \NT $ complex-valued circularly-symmetric Gaussian random matrix with mean $ \uHd $, row covariance $ \mbI_{\NR} $, and column covariance $ \uRT $, i.e., a matrix whose vectorized form is distributed as $ \text{vec}(\uHH) \sim {\cal{CN}}_{\NR\NT} \left( \text{vec}(\uHdH),  \mbI_{\NR} \otimes \uRT \right) $, is denoted herein as $ \uH \sim {\cal{CN}}_{\NR,\NT} \left( \uHd, \mbI_{\NR} \otimes \uRT \right) $, based on the definition from\cite{jin_tit_07}; subscripts $ \cdot_{\text{d}} $ and $ \cdot_{\text{r}} $ identify, respectively, deterministic and random components; subscript $ \cdot_{\text{n}} $ indicates a normalized variable; $ \mathbb{E} \{ \cdot \} $ denotes statistical average; 
$ {\sf \Gamma} (N, \Gamma_1) $ represents the \textit{gamma} distribution with shape parameter $ N $ and scale parameter $\Gamma_1 $; $ \chi_m^2(\delta) $ denotes the noncentral \textit{chi-square} distribution with $ m $ degrees of freedom and noncentrality parameter $ \delta $; $ \chi_m^2 $ denotes the central chi-square distribution with $ m $ degrees of freedom; $ {\sf B}(N, M) $ represents the central \textit{beta} distribution with shape parameters $ N $ and $ M $; $ {\sf B}(N, M, x) $ represents the noncentral {beta} distribution with shape parameters $ N $ and $ M $, and noncentrality $ x $.
\item $ {\hFoo}(\cdot; \cdot; \cdot) $ is the \textit{confluent hypergeometric function}\cite[Eq.~(13.2.2)]{NIST_book_10}; $ (N)_n $ is the Pochhammer symbol, i.e.,  $ (N)_0 = 1 $ and $ (N)_n = N (N + 1) \ldots (N + n - 1) $, $ \forall n \geq 1 $.
\item $ \partial_t^k g(t,z) $ denotes the $ k $th partial derivative w.r.t.~$ t $ of function $ g(t,z) $.
\end{itemize}

\section{Model and Assumptions}
\label{section_system_channel_model}

\subsection{Received Signal and Fading Models}
\label{section_signal_model}

We consider an uncoded point-to-point uplink MIMO spatial multiplexing system over a frequency-flat fading channel\cite[Chs.~3, 7]{tse_book_05}.
There are $ \NT \ge 2 $ and $ \NR \ge \NT $ antenna elements at the transmitter\footnote{For $ \NT = 1 $ and maximal-ratio combining (MRC), we obtained a simple {\sf SNR}  m.g.f.~expression for Rician fading in\cite[Eq.~(36)]{siriteanu_twc_13}.} and receiver, respectively.
For the transmit-symbol vector denoted with
\begin{eqnarray}
\label{equation_transmitted_vector}
\uy =\begin{matrix} ( {y_1} & {y_2} & {\cdots} & {y_{\NT}})  \end{matrix}^{\sf T} \doteq \NT \times 1,
\end{eqnarray}
the stream of complex-valued symbols $ y_i $ from antenna $ i $ is referred to as Stream $ i $. 
Without loss of generality, we consider Stream 1 as the intended stream (i.e., whose symbol is detected, and whose detection performance is analyzed and evaluated), and the remaining 
\begin{eqnarray}
\NI = \NT - 1
\end{eqnarray}
streams, i.e., Streams $ i = 2 : \NT $, as interfering streams.
The number of \textit{degrees of freedom} is
\begin{eqnarray}
\label{equation_DoFs}
N = \NR - \NI= \NR - \NT + 1.
\end{eqnarray}

Then, the received signal vector can be represented as
\begin{eqnarray}
\label{equation_received_signal}
\ur = \sqrt{\frac{{E}_{\text{s}}}{\NT}} \, \uH \uy + \un \doteq \NR \times 1,
\end{eqnarray}
where $ \frac{E_{\text{s}}}{\NT} $ is the energy transmitted per symbol (i.e., per antenna), and 
 $\un \sim {\cal{CN}}_{\NR} (\mzero, N_0 \, \mbI_{\NR})$ is the additive noise.
Then, the per-symbol transmit SNR is
\begin{eqnarray}
\label{equation_Gammas}
\Gamma_{\text{s}} = \frac{{E}_{\text{s}}}{N_0} \frac{1}{\NT}.
\end{eqnarray}
Finally, we assume that the complex-valued channel matrix $\uH \doteq \NR \times \NT $ is Gaussian (more details follow below), has rank $ \NT $, and is perfectly known at the receiver\footnote{{\sf ZF} for imperfectly-known $ \uH $ can be studied, e.g., with the effective-{\sf SNR}  approach we described in\cite{siriteanu_tvt_11}.}.
With its deterministic and random components denoted as $\uHd$ and $\uHr$, respectively, we can write
\begin{eqnarray}
\label{equation channel_matrix_components}
\uH =  \uHd + \uHr = \sqrt{\frac{K}{K+1}} \, \uHdn + \sqrt{\frac{1}{K+1}} \, \uHrn,
\end{eqnarray}
where $ \uHdn $ and $ \uHrn $ are the components of $ \uH $ normalized as
\begin{eqnarray}
\label{equation_Hdn_norm_assumption}
\label{equation_Hrn_norm_assumption}
\| \uHdn \|^2 = \mathbb{E} \{ \| \uHrn \|^2 \}  =  \NR \NT \text{, i.e., }  \mathbb{E} \{ \| \uH \|^2 \} = \NR \NT,
\end{eqnarray}
and $ K $, known as the Rician $ K $-factor, is described by
\begin{eqnarray}
\label{equation K_definition}
K = \frac{\| \uHd \|^2}{\mathbb{E} \{ \| \uHr \|^2 \} } = \frac{ \frac{K}{K+1} \| \uHdn \|^2}{\frac{1}{K+1} \mathbb{E} \{ \| \uHrn \|^2 \} }.
\end{eqnarray}

Then, $ K = 0 $ yields full-Rayleigh fading, i.e., $ |\left[ \uH \right]_{i,j}| $ is Rayleigh distributed $ \forall i, j $, as assumed in\cite{winters_tcom_94},\cite{gore_cl_02},\cite{kiessling_spawc_03}.
Further, the case when $ K \neq 0 $ and in $ \uHdn $ only column $ [\uHdn]_{\bullet,1} $ is nonzero is referred to as Rician--Rayleigh fading, as in\cite{siriteanu_twc_13},\cite{siriteanu_twc_14}.
Finally, herein, the case when $ K \neq 0 $ and each column of $ \uHdn $ has at least one nonzero element is referred to as full-Rician fading.


We assume that $ \uHd $ arises due to LoS propagation between transmitter and receiver.
Then, if the transmitter--receiver distance is much larger than the antenna interelement spacing, $ \uHd $ can be represented as the outer product of the array response vectors for the receiving antenna, $ \ua \doteq \NR \times 1 $, and transmitting antenna, $ \ub \doteq \NT \times 1 $,
i.e.,\cite[Eq.~(7.29), p.~299]{tse_book_05}
\begin{eqnarray}
\label{equation_Hd_outer_product}
\uHd = \ua \ub^{\sf H} = \ua \, \begin{matrix} (b_1^* & b_2^* & \dots & b_{\NT}^*) \end{matrix},
\end{eqnarray}
which reveals that $ \uHd $ has rank $ r = 1 $ and columns given by $ \uh_{\text{d},i} =\ua \, b_i^*, i = 1 : \NT $. 
\begin{remark}
\label{remark_b_norm}
We may assume that $ \| \ua \| = 1 $ if we scale  
$\ub$ according to
\begin{eqnarray}
\label{equation_b_norm_squared}
\| \ub \|^2 = \sum_{i=1}^{\NT} | b_i |^2 =  \sum_{i=1}^{\NT}\underbrace{\| \ua \|^2}_{=1} | b_i |^2  =  \sum_{i=1}^{\NT} \| \uh_{\text{d},i} \|^2 = \| \uHd \|^2  {\buildrel (\ref{equation channel_matrix_components}),(\ref{equation_Hdn_norm_assumption}) \over =} \frac{K}{K+1} \NR \NT.
\end{eqnarray}
\end{remark}

For a tractable analysis, we assume zero row correlation (i.e., receive-antenna correlation) for $ \uH $. 
On the other hand, we assume, as in\cite{gore_cl_02},\cite{kiessling_spawc_03},\cite{siriteanu_twc_13},\cite{siriteanu_twc_14}, that any row of $ \uHrn $ has the same distribution $ {\cal{CN}}_{\NT} (\mzero, \uRT ) $, so that any row of $ \uHr $ has the same distribution $ {\cal{CN}}_{\NT} (\mzero, \uRTK ) $ with
\begin{eqnarray}
\label{equation_RTK_Hr_RT}
\uRTK = \frac{1}{\NR}\mathbb{E} \{ \uHrH \uHr\} = \frac{1}{K + 1 } \frac{1}{\NR}\mathbb{E} \{ \uHrnH \uHrn \} = \frac{1}{K + 1 } \uRT.
\end{eqnarray}
Thus, we can write $ \uHr = \uHw \uRTKsqrt $ with $ \uHw \sim {\cal{CN}}_{\NR,\NT} \left( \mzero, \mbI_{\NR} \otimes \mbI_{\NT} \right) $, 
so that $ \uH = \uHd + \uHr \sim {\cal{CN}}_{\NR,\NT} \left( \uHd, \mbI_{\NR} \otimes \uRTK \right) $.

Matrix $ \uRT $ is determined by antenna interelement spacing and AS, i.e., the `standard deviation' of the power azimuth spectrum\cite[p.~136]{siriteanu_thesis_06}. When the latter is modeled as Laplacian, as recommended by WINNER II\cite{winner_d_1_1_2_v_1_2}, $ \uRT $ can be computed from the AS with\cite[Eqs.~(4-3)--(4-5)]{siriteanu_thesis_06}.

\begin{remark}
WINNER II modeled the measured AS (in degrees) and $ K $ (in dB) as random variables with scenario-dependent lognormal distributions\cite[Table 5.5]{winner_d_1_1_2_v_1_2}\cite[Table~1]{siriteanu_tvt_11}.
Thus, herein, we attempt to evaluate {\sf ZF} performance for AS and $ K $ values relevant to these distributions.
\end{remark}



\subsection{Matrix Partitioning Used in Analysis}

To study Stream-1 detection performance, we shall employ the partitioning 
\begin{eqnarray}
\label{equation_partitioned_H_Hd_Hr_Hdn_Hrn}
\uH = \begin{matrix} ( {\uha} & {\uHb})  \end{matrix} = \begin{matrix} ( \uhda \quad \uHdb \end{matrix} ) + \begin{matrix} ( \uhra \quad \uHrb \end{matrix} ),
\end{eqnarray}
where $ \uha $, $ \uhda $, and $ \uhra $ are $ \NR \times 1 $ vectors, whereas $ \uHb $, $ \uHdb $, and $ \uHrb $ are $ \NR \times \NI $ matrices.
We shall also employ the corresponding partitioning of the column covariance matrix:
\begin{eqnarray}
\label{equation_partitioned_RTK}
\uRTK = \left(
  \begin{array}{cc}
    \RTaa & \RTab \\
    \RTba & \RTbb \\
  \end{array}
  \right) = \left(
  \begin{array}{cc}
     \rTaa  & \rTbaH \\
    \rTba & \RTbb \\
  \end{array}
  \right).
\end{eqnarray}

\begin{remark}
\label{remark_special_case_btilde}
Herein, we consider full-Rician fading with $ r = \text{rank}(\uHd) = \text{rank}(\uHdb) = 1 $, whereas in\cite{siriteanu_twc_13},\cite{siriteanu_twc_14} we considered its special case of Rician--Rayleigh fading, i.e., $ \text{rank}(\uHd) = 1 $, but $ \text{rank}(\uHdb) = 0 $.
Thus, the results obtained herein specialize to those in\cite{siriteanu_twc_13},\cite{siriteanu_twc_14} when we reduce to $ \mzero $ the vector formed with the last $ \NI = \NT - 1 $ elements of $ \ub $, i.e., the vector
\begin{eqnarray}
\label{equation_btilde}
\ubtilde = \begin{matrix}(b_2 & \dots &  b_{\NT})^{\sf T} \end{matrix}.
\end{eqnarray}
\end{remark}

\section{Exact Analysis of {\sf ZF SNR}}
\label{section_MGF_unconditioned_SNR}

\subsection{{\sf ZF} {\sf SNR}  as Hermitian Form}
\label{section_ZF_SNR_Hermitian_Schur}

Given~$\uH$, {\sf ZF} for the signal from~(\ref{equation_received_signal}) refers to symbol detection based on the operation
\begin{eqnarray}
\label{equation ZF_for_perfect_CSI}
\sqrt{\frac{ \NT }{ {E}_{\text{s}} }} \left[ \uH^{\sf H} \uH \right]^{-1} \uH^{\sf H} \, \ur
= \uy +\frac{ 1 }{  \sqrt{ { \Gamma_{\text{s}} }  }} \left[ \uH^{\sf H} \uH \right]^{-1} \uH^{\sf H} \frac{\un}{\sqrt{N_0}}.
\end{eqnarray}
Based on~(\ref{equation ZF_for_perfect_CSI}) and\cite{kiessling_spawc_03},\cite{siriteanu_twc_13}, the {\sf SNR}  for Stream 1 can be written as the Hermitian form below:
\begin{eqnarray}
\label{equation_gammak_perfect_CSI}
\label{equation_SNR_Hermitian}
\gamma_1 = \frac{ \Gamma_{\text{s}} }{\left[(\uHH \uH )^{-1}\right]_{1,1}} = \Gamma_{\text{s}} \uhaH \underbrace{ \big[ \mbI_{\NR} - \uHb (\uHbH \uHb)^{-1} \uHbH \big]}_{=\uQ_2} \uha,
\end{eqnarray}
where $ \uQ_2 \doteq \NR \times \NR $ is idempotent and of rank $ N $.

\begin{remark}
\label{remark_announce_transf}
The following transformations do not change the {\sf ZF} {\sf SNR}  in~(\ref{equation_gammak_perfect_CSI}):
\begin{itemize}
\item Row transformations of $ \uH $ with unitary matrices, because they do not change $ \uHH \uH $.
\item Column transformations of $ \uHb $ with nonsingular matrices, because they do not change $ \uQ_2  $.
\end{itemize}
Several such transformations, shown below, help derive the exact {\sf SNR} distribution.
\end{remark}

\subsection{Row Transformation $ \uF = {\mb{V}}\uH $ That Zeroes Rows $ {[{\uFd}]_{i,\bullet}} $, $i=2:\NR$}
\label{section_row_trans}

If we make the substitution $ \uH = {\mb{V}}^{\sf H} \uF $, with unitary $  {\mb{V}} \doteq \NR \times \NR $, in~(\ref{equation_SNR_Hermitian}) and partition according to~(\ref{equation_partitioned_H_Hd_Hr_Hdn_Hrn}) the matrix
\begin{eqnarray}
\label{equation_partitioned_F_Fd_Fr_Fdn_Frn}
\uF & = & {\mb{V}}\uH \doteq \NR \times \NT \nonumber \\
& = & \begin{matrix} ( {\ufa} & {\uFb})  \end{matrix} = \begin{matrix} ( \ufda \quad \uFdb \end{matrix} ) + \begin{matrix} ( \ufra \quad \uFrb \end{matrix} ),
\end{eqnarray}
the {\sf ZF SNR} Hermitian form in~(\ref{equation_SNR_Hermitian}) becomes
\begin{eqnarray}
\label{equation_SNR_Hermitian_F}
\gamma_1 = \Gamma_{\text{s}} \ufaH \uQ_2 \ufa,
\end{eqnarray}
with
\begin{eqnarray}
\label{equation_Q2_H2_F2}
\uQ_2  = \mbI_{\NR} - \uFb (\uFbH \uFb)^{-1} \uFbH.
\end{eqnarray} 


Choosing the first row of the unitary matrix $ \uV $ as $  [\uV]_{1,\bullet} = \uaH $, we conveniently obtain
\begin{eqnarray}
\label{equation_Fd_first_row}
&& {[{\uFd}]_{1,\bullet}}  { \,{\buildrel {(\ref{equation_Hd_outer_product})} \over =}\, }  ([\uV]_{1,\bullet} \, \ua) \ub^{\sf H}   = \|\ua\|^2 \ub^{\sf H} =  \ub^{\sf H}, \nonumber \\
\label{equation_Fd_remaining_rows}
&& {[{\uFd}]_{i,\bullet}}  { \,{\buildrel {(\ref{equation_Hd_outer_product})} \over =}\, }  \underbrace{([\uV]_{i,\bullet} \, \ua)}_{=0} \ub^{\sf H} = \mzero, \;\; i = 2 : \NR, \nonumber \\
\label{equation_F_columns}
&& \text{i.e., } {[{\uFd}]_{\bullet,j}} = \uf_{\text{d},j} = \begin{matrix} \, (b_j^* & 0 & \dots & 0) \end{matrix}^{\sf T}, \, j = 1 :  \NT. \quad \quad
\end{eqnarray}


\begin{theorem}
\label{theorem_gamma1_cond_on_Q2_distribution}
The m.g.f.~of the {\sf SNR}  conditioned on $ \uQ_2 $ can be written, simply, as
\begin{eqnarray}
\label{equation_gamma1_mgf_cond_Q2_11}
M_{\gamma_1|\uQ_2}(s) = \mathbb{E}_{{\gamma_1}} \{ e^{s \gamma_1} | \uQ_2 \} = \frac{1}{(1-\Gamma_1 s)^N}  \exp \big\{  f_1(s) [\uQ_2]_{1,1} \big\},
\end{eqnarray}
with scalar $ \Gamma_1  $ and function $ f_1(s) $  defined in the proof below.
\end{theorem}

\begin{IEEEproof}

Because the column covariance of $ \uF = {\mb{V}}\uH $ is the same as that of $ \uH $, i.e., $ \uRTK $, partitioned as in~(\ref{equation_partitioned_RTK}), and because $ \ufa \doteq \NR \times 1 $ and $ \uFb \doteq \NR \times \NI  $ from the partitioning of $ \uF $ in~(\ref{equation_partitioned_F_Fd_Fr_Fdn_Frn}) are jointly Gaussian, the distribution of $ \ufa $ given $ \uFb $ is given by\cite[Appendix]{kiessling_spawc_03},\cite[Eqs.~(12)--(16)]{siriteanu_twc_13}
\begin{eqnarray}
\label{equation_distribution_f1_given_F2}
\ufa | \uFb \sim {\cal{CN}}_{\NR} \bigg( \underbrace{(\ufda - \uFdb \ur_{2,1})}_{ =\umu \doteq \NR \times 1} + \uFb \ur_{2,1}, \left( \RTinvaa \right)^{-1} \mbI_{\NR} \! \bigg),
\end{eqnarray}
with
\begin{eqnarray}
\label{equation_r21}
\ur_{2,1} & = & \RTbbinva \rTba \doteq \NI \times 1, \\
\label{equation_RTKinv11}
\left( \RTinvaa \right)^{-1} \! \! \! \! \! \!  & =& \rTaa - \rTbaH \, \RTbbinva \, \rTba.
\end{eqnarray}

Then, it can be shown by
substituting~(\ref{equation_distribution_f1_given_F2}) into~(\ref{equation_SNR_Hermitian_F}) and further manipulating as in\cite{kiessling_spawc_03},\cite{siriteanu_twc_13}, that the {\sf SNR}  conditioned on $ \uQ_2 $ from~(\ref{equation_SNR_Hermitian_F}) can be written as the Hermitian form
\begin{eqnarray}
\label{equation_gamma1_cond_Q2_f1_tilde}
\gamma_1 | \uQ_2  & = &  \Gamma_1 \ufatildeH \uQ_2 \ufatilde, \quad \text{with} \\
\label{equation_Gamma1}
\Gamma_1 & = & \frac{ { \Gamma_{\text{s}} } }{\RTinvaa}, \\
\label{equation_distribution_f1tilde}
\ufatilde & \sim & {\cal{CN}}_{\NR} \left( {\sqrt{\RTinvaa}} \, \umu, \mbI_{\NR} \right), \\
\label{equation_umu_simple_form}
\umu & { \,{\buildrel {(\ref{equation_distribution_f1_given_F2})} \over =}\, } & \ufda - \uFdb \ur_{2,1} { \,{\buildrel {(\ref{equation_F_columns})} \over =}\, }  \begin{matrix} (b_1^* - \ubtilde^{\sf H} \ur_{2,1} & 0 & \dots & 0)^{\sf T} \end{matrix}
= \begin{matrix} (\mu_1 & 0 & \dots & 0)^{\sf T} \end{matrix},
\end{eqnarray}
i.e., row transformation $ \uF = {\mb{V}}\uH $ yielded a single nonzero-mean element in $ \ufatilde $, which simplifies the ensuing analysis.
%

The Hermitian form in $ \ufatilde $ from~(\ref{equation_gamma1_cond_Q2_f1_tilde}) helps cast the m.g.f.~of the {\sf SNR}  given $ \uQ_2 $ as\cite[Eq.~(20)]{siriteanu_twc_13}
\begin{eqnarray}
\label{equation_gamma1_mgf_cond_Q2_Turin_nu}
 M_{\gamma_1|\uQ_2}(s) 
= \frac{\exp \big\{ -  x_1 \unu^{\sf H} \left[ \mbI_{\NR} - \left( \mbI_{\NR} - \Gamma_1 s  \uQ_2 \right)^{-1} \right] \unu \big\}}{ \det \left( \mbI_{\NR} - \Gamma_1 s  \uQ_2 \right)},
\end{eqnarray}
with
\begin{eqnarray}
\label{equation_x1_scalar}
&  x_1 = \RTinvaa \| \umu \|^2 = \RTinvaa | \mu_1 |^2, \\
\label{equation_unu_vector}
&  \unu = \frac{\umu}{ \mu_1 } = \begin{matrix} (1 & 0 & \dots & 0)^{\sf T} \end{matrix}, \quad \quad \quad \quad \\
\label{equation_derivation_Herm_form_matrix}
&  \mbI_{\NR} - \left( \mbI_{\NR} - \Gamma_1 s  \uQ_2 \right)^{-1} = - \frac{\Gamma_1 s  }{1 - \Gamma_1 s  } \uQ_2.
\end{eqnarray}
Above,~(\ref{equation_derivation_Herm_form_matrix}) follows by using the eigendecomposition of $ \uQ_2 $.
The desired m.g.f.~expression in~(\ref{equation_gamma1_mgf_cond_Q2_11}) follows by substituting~(\ref{equation_derivation_Herm_form_matrix}) into~(\ref{equation_gamma1_mgf_cond_Q2_Turin_nu}) and defining $ f_1(s)  = \frac{\Gamma_1 s  }{1 - \Gamma_1 s  } \, x_1 $.

\end{IEEEproof}

\subsection{Partial Column Transformations That Help Rewrite $ [\uQ_2]_{1,1} $ Conveniently}

\subsubsection{Unitary Transformation $ \uEb = \uFb \uVtilde $ That Zeroes Elements $ [\uEdb]_{1,j} $, $ j = 2 : \NI $}
\label{section_col_trans}

Making the substitution $ \uF_2 = \uE_2 \uVtildeH $, with unitary $\uVtilde \doteq \NI \times \NI $, in~(\ref{equation_Q2_H2_F2}) yields
\begin{eqnarray}
\label{equation_Q2_F2_E2}
\uQ_2  = \mbI_{\NR} - \uEb (\uEbH \uEb)^{-1} \uEbH.
\end{eqnarray} 
Based on~(\ref{equation_partitioned_F_Fd_Fr_Fdn_Frn}), we can write
\begin{eqnarray}
\label{equation_E2}
\uEb = \uFb \uVtilde = \uFdb \uVtilde + \uFrb \uVtilde = \uEdb + \uErb \doteq \NR \times \NI.
\end{eqnarray} 
Setting $ [\uVtilde]_{\bullet,1} = {\ubtilde}/{ \| \ubtilde \|} $ simplifies the ensuing {\sf SNR}  analysis as it zeroes $ [\uEdb]_{1,j} $, $ j = 2 : \NI $:
\begin{eqnarray}
\label{equation_Ed2}
\uEdb = \uFdb \uVtilde { \,{\buildrel {(\ref{equation_F_columns})} \over =}\, }  \left(
  \begin{array}{c}
      \ubtilde^{\sf H} \\
    \mzero \\
  \end{array}
  \right) \left(
  \begin{array}{cc}
      \frac{\ubtilde}{ \| \ubtilde \|} & [\uVtilde]_{\bullet,2} \cdots [\uVtilde]_{\bullet,\NI} \\
  \end{array}
  \right) = \| \ubtilde \| \left(
  \begin{array}{cc}
     1 & \mzero \\
    \mzero & \mzero \\
  \end{array}
  \right).
\end{eqnarray}

\subsubsection{Nonsingular Transformation That Decorrelates the Columns of $ \uEb $}
\label{section_H2_nonzero_mean_covariance}


For the column correlation of $ \uErb $ from~(\ref{equation_E2}), i.e., for
\begin{eqnarray}
\label{equation_E2_correlation}
\frac{1}{\NR}\mathbb{E} \{ \uErbH \uErb \} = \frac{1}{\NR}\mathbb{E} \{ (\uFrb \uVtilde)^{\sf H} (\uFrb \uVtilde) \}  
{ \,{\buildrel {(\ref{equation_partitioned_RTK})} \over =}\, } \uVtilde^{\sf H} \RTbb \uVtilde,
\end{eqnarray} 
let us consider the Cholesky decomposition\cite[Sec.~5.6]{gentle_book_07} 
\begin{eqnarray}
\label{equation_Choleski_for_E2_Correlation}
\uVtilde^{\sf H} \RTbb \uVtilde = \uA \uAH,
\end{eqnarray}
where $ \uA \doteq \NI \times \NI $ is upper triangular with real-valued and positive diagonal elements. 

Then, considering matrix $ \uEwb \sim {\cal{CN}}_{\NR,\NI} \left( \mzero, \mbI_{\NR} \otimes \mbI_{\NI} \right) $, we can write~(\ref{equation_E2}) based on~(\ref{equation_Choleski_for_E2_Correlation}) and~(\ref{equation_E2_correlation}) as
\begin{eqnarray}
\label{equation_E2_components}
\uEb = \uEdb + \uEwb \uAH = \left( \uEdb \uAinvH  + \uEwb \right) \uAH.
\end{eqnarray}  
Thus, by transforming the columns of $ \uEb $ with $ \uAinvH $, we obtain
\begin{eqnarray}
\label{equation_G2_definition}
\uG_2 =  \uEb \uAinvH =  \uEdb \uAinvH  + \uEwb  \doteq \NR \times \NI,
\end{eqnarray}
whose mean can be written, based on~(\ref{equation_Ed2}) and the fact that $ \uAinvH $ is lower triangular, as
\begin{eqnarray}
\label{equation_Gdb}
\uGdb = \uEdb \uAinvH = \| \ubtilde \| [\uAinvH]_{1,1}  \left(
  \begin{array}{cc}
     1 & \mzero \\
    \mzero & \mzero \\
  \end{array}
  \right).
\end{eqnarray}
Using~(\ref{equation_Choleski_for_E2_Correlation}), the properties of $ \uA $, and the choice $ [\uVtilde]_{\bullet,1} = {\ubtilde}/{ \| \ubtilde \|} $, the squared norm of $ \uGdb $ can be written as
\begin{eqnarray}
\label{equation_x2_Gd2_norm}
x_2 = \| \uGdb \|^2 = \ubtilde^{\sf H} \RTbbinva \ubtilde.
\end{eqnarray}

\begin{remark}
\label{remark_special_case_x2}
For Rician--Rayleigh fading, Remark~\ref{remark_special_case_btilde} revealed that $ \ubtilde = \mzero $, which by~(\ref{equation_x2_Gd2_norm}) implies $ x_2 = 0 $.
On the other hand, for full-Rayleigh fading,~(\ref{equation_x1_scalar}) implies that also $ x_1 = 0 $.
\end{remark}

Thus, column transformation~(\ref{equation_G2_definition}) yielded $ \uGb $ with uncorrelated columns and mean given by
\begin{eqnarray}
\label{equation_Gd2_elements}
[\uGdb]_{i,j} = \begin{cases} \sqrt{x_2} & , \text{(i.e., real-valued) for } i=j=1, \\
0 & \text{, otherwise.} \end{cases}
\end{eqnarray}
Substituting $ \uEb = \uG_2 \uAH  $ into~(\ref{equation_Q2_F2_E2}) yields
\begin{eqnarray}
\label{equation_Q2_E2_G2}
\uQ_2 = \mbI_{\NR} - \uGb (\uGbH \uGb)^{-1} \uGbH.
\end{eqnarray}
The simple statistics of $ \uGb $ (vs. $ \uFb $) help simplify our {\sf SNR} distribution analysis, as shown below.


\subsubsection{QR Decomposition}
\label{section_G2_QR}

Finally, by substituting in~(\ref{equation_Q2_E2_G2}) the QR decomposition\cite[Sec.~5.7]{gentle_book_07}
\begin{eqnarray}
\label{equation_G2_QR_decomposition}
 \uGb = \uUb \uT_2,
\end{eqnarray}
where $ \uUb \doteq \NR \times \NI $ satisfies $ \uUb^{\sf H}\uUb 
= \mbI_{\NI} $, and $ \uT_2 \doteq \NI \times \NI $ is upper triangular with real-valued and positive diagonal elements, we can write $ \uQ_2 $ simply as 
\begin{eqnarray}
\label{equation_Q2_G2_U2}
\uQ_2  = \mbI_{\NR} - \uUb \uT_2 (\uT_2^{\sf H} \uT_2)^{-1} \uT_2^{\sf H} \uUb^{\sf H}
\label{equation_Q2_after_QR}
= \mbI_{\NR} - \uUb \uUb^{\sf H}.
\end{eqnarray}
This helps write $ [\uQ_2]_{1,1}  $ for the m.g.f.~in~(\ref{equation_gamma1_mgf_cond_Q2_11}) solely in terms of the first row of $ \uUb $ as 
\begin{eqnarray}
\label{equation_Q211_as_product}
[\uQ_2]_{1,1}  & = & 1 -  [\uUb]_{1,\bullet} \, ([\uUb]_{1,\bullet})^{\sf H} = 1 - (|[\uUb]_{1,1}|^2  + |[\uUb]_{1,2}|^2 + \dots + |[\uUb]_{1,\NI}|^2) \nonumber \\
& = & \underbrace{(1 - |[\uUb]_{1,1}|^2)}_{=\beta_1}  \bigg( \underbrace{1 - \frac{ |[\uUb]_{1,2}|^2 + \dots + |[\uUb]_{1,\NI}|^2}{1 - |[\uUb]_{1,1}|^2}}_{ = {{\beta_2}}} \bigg).
\end{eqnarray}

\subsection{Principal Analysis Result: Exact M.G.F.~Expression of the Unconditioned {\sf SNR} }

The above transformations have helped write the conditioned-{\sf SNR}  m.g.f.~from~(\ref{equation_gamma1_mgf_cond_Q2_11}) as
\begin{eqnarray}
\label{equation_mgf_gamma1_cond_rho_U211}
M_{\gamma_1}(s \mid \beta_1, {{\beta_2}}) =\frac{1}{(1-\Gamma_1 s)^N} 
\exp \{ {f}_1(s) \beta_1 {{\beta_2}} \}.
\end{eqnarray}
In order to express the unconditioned-{\sf SNR}  m.g.f., we need to average~(\ref{equation_mgf_gamma1_cond_rho_U211}) over the distributions of $ \beta_1 $ and $ {{\beta_2}} $, which are elucidated in the following two lemmas.




\begin{lemma}
\label{lemma_U211_distribution}
Random variable $ \beta_1  $ from~(\ref{equation_Q211_as_product}) is distributed as
\begin{eqnarray}
\label{equation_one_minus_U211_distribution}
\beta_1 \sim {\sf B}(\NR - 1, 1, {x_{2}}).
\end{eqnarray}
\end{lemma}

\begin{IEEEproof}
See Appendix~\ref{lemma_2}.
\end{IEEEproof}

\begin{lemma} 
\label{lemma_rho_distribution}
Random variable $ {\beta_2} $ from~(\ref{equation_Q211_as_product}) is distributed as
\begin{eqnarray}
\label{equation_rho_distribution_beta}
{\beta_2} \sim {\sf B}(N, \NI-1),
\end{eqnarray}
i.e., has m.g.f.\cite[Eq.~(30)]{siriteanu_twc_13} 
\begin{eqnarray}
\label{equation_mgf_rho}
M_{{\beta_2}}(s) = {\hFoo(N; \NR -1; s)},
\end{eqnarray}
and is independent of $ \beta_1 $.
\end{lemma}
\begin{IEEEproof}
See Appendix~\ref{lemma_1}.
\end{IEEEproof}

\begin{theorem}
\label{theorem_gamma1_mgf_main_result}
The m.g.f.~of the unconditioned {\sf ZF}  {\sf SNR}  under full-Rician fading with $ r = 1 $ is
\begin{eqnarray}
\label{equation_mgf_gamma1_uncond_main_result}
M_{\gamma_1}(s; {x}_1, x_{2}) & = & \frac{1}{(1-\Gamma_1 s)^N} \sum_{{n_{2}}=0}^\infty \frac{ e^{-{x_{2}}} {x_{2}^{n_{2}}}}{{n_{2}}!}  {\hFoo}\!\left(N; n_{2} + \NR;  \frac{ \Gamma_1 s}{1- \Gamma_1 s} {x}_1 \right). \quad \quad
\end{eqnarray}
\end{theorem}

\begin{IEEEproof}
Due to limited space, we only outline the proof: it follows by successively averaging the m.g.f.~of the conditioned {\sf SNR}  in~(\ref{equation_mgf_gamma1_cond_rho_U211}) over the distributions of the independent random variables  $ {{\beta_1}} $ and $ {{\beta_2}} $, and by exploiting~(\ref{equation_mgf_rho}),~(\ref{equation_1F1_series}), and~(\ref{equation_U211_moments}).
\end{IEEEproof}

\subsection{Effects of Channel Matrix Statistics on {\sf SNR} Statistics}
\label{section_SNR_statistics}

For Rician--Rayleigh fading (i.e., for $ x_2 = 0 $), the {\sf SNR}  m.g.f.~from~(\ref{equation_mgf_gamma1_uncond_main_result}) reduces to\cite[Eq.~(31)]{siriteanu_twc_13}
\begin{eqnarray}
\label{equation_mgf_gamma1_Rice_Ray}
M_{\gamma_1}(s; {x}_1)  =  \frac{1}{(1-\Gamma_1 s)^N} {\hFoo}\!\left(N; \NR;  \frac{ \Gamma_1 s}{1- \Gamma_1 s} {x}_1 \right).
\end{eqnarray}
Then, for $ \gamma_1 $, the first two moments, variance $ \mathbb{V} \{ \gamma_1 \} = \mathbb{E} \{ \gamma_1^2 \} - \left( \mathbb{E} \{ \gamma_1 \} \right)^2 $, and amount of fading $ \mathbb{A} \{ \gamma_1 \} = \mathbb{V} \{ \gamma_1 \}/\left( \mathbb{E} \{ \gamma_1 \} \right)^2 $, i.e., {\sf SNR}  statistics, have been expressed in\cite[Table~I]{siriteanu_twc_13}.
Because the {\sf SNR}  m.g.f.~for full-Rician fading from~(\ref{equation_mgf_gamma1_uncond_main_result}) is a weighted infinite series of {\sf SNR}  m.g.f.'s for Rician--Rayleigh fading from~(\ref{equation_mgf_gamma1_Rice_Ray}) with $ \NR $ replaced with $ \NR + n_2 $, expressing $ \mathbb{E} \{ \gamma_1 \} $ and $ \mathbb{E} \{ \gamma_1^2 \} $ for the former from those for the latter from\cite[Table~I]{siriteanu_twc_13} is trivial.
Expressing $ \mathbb{V} \{ \gamma_1 \} $ and $ \mathbb{A} \{ \gamma_1 \} $ based on~(\ref{equation_mgf_gamma1_uncond_main_result}) and\cite[Table~I]{siriteanu_twc_13} is not trivial.

On the one hand, the effect of $ x_2 $ on {\sf SNR}  statistics is not readily discernible from~(\ref{equation_mgf_gamma1_uncond_main_result}) and \cite[Table~I]{siriteanu_twc_13}. 
On the other hand,~(\ref{equation_mgf_gamma1_uncond_main_result}) and \cite[Table~I]{siriteanu_twc_13} reveal that $ \mathbb{E} \{ \gamma_1 \} $ increases with $ N $ from~(\ref{equation_DoFs}), $\Gamma_1 $ from~(\ref{equation_Gamma1}), and $x_1 $ from~(\ref{equation_x1_scalar}).
Further, note that it can be shown that $x_1 \propto \| \uhda - \uHdb \ur_{2,1} \| $. Thus, the performance of {\sf ZF} for full-Rician fading with $ r = 1 $ is worst when the channel matrix statistics satisfy condition $  \uhda = \uHdb \ur_{2,1} $, and it improves with increasing $ \| \uhda - \uHdb \ur_{2,1} \|  $.
In\cite{siriteanu_twc_SC_14}, where we studied full-Rician fading irrespective of $ r $, we had noticed (e.g., by comparing\cite[Figs.~1, 2]{siriteanu_twc_SC_14}) that {\sf ZF} performed worst for $ \uhda = \uHdb \ur_{2,1} $.

\begin{remark}
\label{remark_Rayleigh_approx}
Note that $ \forall x_2 $, if $ \uhda = \uHdb \ur_{2,1} $, i.e., $ x_1 = 0 $, then the m.g.f.~in~(\ref{equation_mgf_gamma1_uncond_main_result}) reduces to the gamma m.g.f.~$ M_{\gamma_1}(s)  =  (1-\Gamma_1 s)^{-N} $.
On the other hand, the gamma distribution with m.g.f.~$ M(s)= (1 - s \widehat{\Gamma}_1)^{-N} $ and $ \widehat{\Gamma}_1 $ obtained as in~(\ref{equation_Gamma1}) from $ \widehat{\mbR}_{\text{T},K} = \mbR_{\text{T},K} + \frac{1}{\NR} \uHdH \uHd $, has previously been employed to approximate the actual {\sf ZF SNR} distribution for Rician fading, irrespective of $ r $ --- see\cite{siriteanu_tvt_11},\cite{matthaiou_tcomm_10} and references therein. 
Interestingly, condition $ \uhda = \uHdb \ur_{2,1} $ yields  $ \Gamma_1 = \widehat{\Gamma}_1 $, rendering the approximation exact --- see\cite[Corollary~4]{siriteanu_twc_SC_14}.
\end{remark}

The above have yielded the following insights.

\begin{remark}
\label{remark_SNR_statistics}
For {\sf ZF} under full-Rician fading with $ r=1 $, condition $ \uhda = \uHdb \ur_{2,1} $ yields: 1) worst performance; 2) full accuracy for the gamma distribution previously employed to approximate the {\sf SNR} distribution.
\end{remark}

\section{Exact Infinite Series Expressions for  {\sf ZF} Performance Measures}
\label{section_infinite_series_for_all}

\subsection{Infinite Series Expansion of $ {\hFoo}(\cdot; \cdot; \sigma) $ Around $ \sigma_0 = 0 $}

Using the well-known infinite series expansion around $ \sigma_0 = 0 $\cite[Eq.~(30)]{siriteanu_twc_13}
\begin{eqnarray}
\label{equation_1F1_series}
{\hFoo}(N; \NR; \sigma) = \sum_{n = 0}^{\infty} { \frac{\left( N \right)_n}{\left( \NR \right)_n} \frac{\sigma^n}{n!} },
\end{eqnarray}
the {\sf SNR}  m.g.f.~from~(\ref{equation_mgf_gamma1_Rice_Ray}) for Rician--Rayleigh fading can also be written as\cite[Eq.~(37)]{siriteanu_twc_13}
\begin{eqnarray}
\label{equation_mgf_gamma1_Rice_Ray_sum} 
M_{\gamma_1}(s; {x}_1) \!\!\! \!\!\!  & = & \!\!\! \!\!\!  \sum_{n_{1}=0}^{\infty}  \frac{(N)_{n_{1}}}{(\NR)_{n_{1}}}  \frac{{x}_1^{n_{1}}}{n_{1}!}  {  \sum_{{m_1} = 0}^{n_{1}} \!\!\! \binom{n_{1}}{{m_1}} (-1)^{m_1} \!\!\!  \underbrace{\frac{1}{(1 - s \Gamma_1)^{N + n_{1} - {m_1}}}}_{ =M_{n_{1},{m_1}}(s)}  }, \quad
\end{eqnarray}
where $ M_{n_{1},{m_1}}(s) $ is the m.g.f.~of a random variable distributed as $ {\sf \Gamma}(N + n_{1} - m, \Gamma_1) $.

Theoretically,~(\ref{equation_1F1_series}) converges $ \forall \sigma $. 
Nevertheless, the computation of~(\ref{equation_1F1_series}) by truncation incurs inherent numerical convergence difficulties with increasing $ \sigma $\cite{siriteanu_twc_13}.
Consequently, the computation of ensuing measures, e.g., the {\sf ZF} {\sf SNR}  p.d.f., becomes nontrivial at realistic values of $ K $, as revealed in\cite{siriteanu_twc_13},\cite{siriteanu_twc_14}.
Similar difficulties arise also for the case studied herein, i.e., full-Rician fading with $ r = 1 $, upon infinite series expansion of $ {\hFoo}(\cdot; \cdot; \sigma) $ in the {\sf SNR}  m.g.f.~expression from~(\ref{equation_mgf_gamma1_uncond_main_result}), as discussed below.


\subsection{Exact Double-Infinite Series for M.G.F., P.D.F., and Performance Measures}
\label{section_ZF_mgf_infinite_series}

By substituting~(\ref{equation_1F1_series}) into~(\ref{equation_mgf_gamma1_uncond_main_result}) and proceeding as for~(\ref{equation_mgf_gamma1_Rice_Ray_sum}), the {\sf SNR}   m.g.f.~becomes
\begin{eqnarray}
\label{equation_mgf_gamma1_double_series_Gamma_mixture}
M_{\gamma_1}(s; {x}_1, x_{2}) 
= e^{-{x_{2}}}  \sum_{n_{1}=0}^{\infty} \sum_{{n_{2}}=0}^\infty \frac{(N)_{n_{1}}}{(n_{2} + \NR)_{n_{1}}}  \frac{{x}_1^{n_{1}}}{n_{1}!} \frac{ {x_{2}^{n_{2}}}}{{n_{2}}!} \underbrace{  \sum_{{m_1} = 0}^{n_{1}} \binom{n_{1}}{{m_1}} (-1)^{m_1} M_{n_{1},{m_1}}(s)}_{ =M_{n_{1}}(s)}. \quad
\end{eqnarray}
Using the m.g.f.--p.d.f.~Laplace-transform pair corresponding to $ {\sf \Gamma}(N + n_{1} - m, \Gamma_1) $, i.e.,
\begin{eqnarray}
\label{equation_Gamma_mgf_pdf_Laplace}
M_{n_{1},{m_1}}(s) & = &  \frac{1}{(1 - s \Gamma_1)^{N + n_{1} - {m_1}}}, \\ 
p_{n_{1},{m_1}} (t) & = & \frac{t^{(N + n_{1} - {m_1}) - 1} e^{-t/\Gamma_1}}{[(N + n_{1} - {m_1}) - 1]! \,  \Gamma_1^{N + n_{1} - {m_1}}},
\end{eqnarray}
the {\sf ZF} {\sf SNR} p.d.f.~corresponding to~(\ref{equation_mgf_gamma1_double_series_Gamma_mixture}) can be written, analogously, as\footnote{An alternate p.d.f.~expression, for real-valued $ \uH $, appears in\cite[Eqs.~(18), (31)]{siriteanu_spl_16}.}:
\begin{eqnarray}
\label{equation_gamma1_pdf_double_series_Gamma_mixture}
p_{\gamma_1} (t; {x}_1, x_{2}) = e^{-{x_{2}}}  \sum_{n_{1}=0}^{\infty} \sum_{{n_{2}}=0}^\infty \frac{(N)_{n_{1}}}{(n_{2} + \NR)_{n_{1}}}  \frac{{x}_1^{n_{1}}}{n_{1}!} \frac{ {x_{2}^{n_{2}}}}{{n_{2}}!} 
\underbrace{ \sum_{{m_1} = 0}^{n_{1}} \binom{n_{1}}{{m_1}} (-1)^{m_1} p_{n_{1},{m_1}} (t)}_{ = p_{n_{1}} (t)}.
\end{eqnarray}

%

By integrating~(\ref{equation_gamma1_pdf_double_series_Gamma_mixture}), the Stream-1 outage probability at threshold {\sf SNR}  $ \tau $ and the ergodic capacity (i.e., rate) are exactly characterized by analogous infinite series, i.e.,
\begin{eqnarray}
\label{equation_Po_definition}
P_{\text{o}}({x}_1, x_{2} ) & = & \int_{0}^{\tau} p_{\gamma_1} (t; {x}_1, x_{2}) \,\mathrm{d} t \\
\label{equation_Po_double_series_Gamma_mixture}
& = & e^{-{x_{2}}}  \sum_{n_{1}=0}^{\infty} \sum_{{n_{2}}=0}^\infty \frac{(N)_{n_{1}}}{(n_{2} + \NR)_{n_{1}}}  \frac{{x}_1^{n_{1}}}{n_{1}!} \frac{ {x_{2}^{n_{2}}}}{{n_{2}}!} \underbrace{ \sum_{{m_1} = 0}^{n_{1}} \binom{n_{1}}{{m_1}} (-1)^{m_1}  P_{\text{o},n_{1},{m_1}}}_{=P_{\text{o},n_{1}}}, \\
\label{equation_ergodic_capacity_definition}
C ({x}_1, x_{2}) & = & \frac{1}{\ln 2}  \int_{0}^{\infty} \ln (1 + t) p_{\gamma_1} (t; {x}_1, x_{2}) \,\mathrm{d} t \\
\label{equation_capacity_double_series_Gamma_mixture}
& = & e^{-{x_{2}}}  \sum_{n_{1}=0}^{\infty} \sum_{{n_{2}}=0}^\infty \frac{(N)_{n_{1}}}{(n_{2} + \NR)_{n_{1}}}  \frac{{x}_1^{n_{1}}}{n_{1}!} \frac{ {x_{2}^{n_{2}}}}{{n_{2}}!} \underbrace{  \sum_{{m_1} = 0}^{n_{1}} \binom{n_{1}}{{m_1}} (-1)^{m_1}  C_{n_{1},{m_1}}}_{ =C_{n_1}},
\end{eqnarray} 
where\footnote{ $ \gamma (k, x) = \int_{0}^{x} t^{k - 1} e^{-t} \,\mathrm{d} t $ is the \textit{incomplete gamma function}\cite[p.~174]{NIST_book_10}. Integral~(\ref{equation_Capacity_Cn1m}) is expressed in\cite[Eq.~(73)]{siriteanu_twc_13}.}
\begin{eqnarray}
\label{equation_Po_triple_series_term_n1m}
\!\!\!\! P_{\text{o},n_{1},{m_1}} \!\!\!\! \!\! & = & \!\!\!\! \int_{0}^{\tau} p_{n_{1},{m_1}} (t) \,\mathrm{d} t = \frac{\gamma \left( N + n_{1} - {m_1},  \tau/\Gamma_1 \right)}{[(N + n_{1} - {m_1}) - 1]!}, \\
\label{equation_Capacity_Cn1m}
\!\!\!\! C_{n_1,{m_1}} \!\!\!\! \!\!
& = & \!\!\!\! \frac{1}{\ln 2} \int_{0}^{\infty} \ln (1 + t) \, p_{n_{1},{m_1}} (t)  \,\mathrm{d} t.
\end{eqnarray}
Finally, the approach in\cite[Section~V.A]{siriteanu_twc_13} can help express also the average error probability as an infinite series analogous to~(\ref{equation_Po_double_series_Gamma_mixture}) and~(\ref{equation_capacity_double_series_Gamma_mixture}).

On the other hand, the previously employed approximating gamma distribution for the {\sf ZF SNR} mentioned in Remark~\ref{remark_Rayleigh_approx} yields simple performance measures expressions similar to~(\ref{equation_Po_triple_series_term_n1m}) and~(\ref{equation_Capacity_Cn1m}).

\subsection{Generic Single-Infinite Series for M.G.F., P.D.F., and Performance Measures}
\label{section_generic_double_series}

Because~(\ref{equation_mgf_gamma1_double_series_Gamma_mixture}),~(\ref{equation_gamma1_pdf_double_series_Gamma_mixture}),~(\ref{equation_Po_double_series_Gamma_mixture}), and~(\ref{equation_capacity_double_series_Gamma_mixture}) are analogous, we may represent them as the generic double infinite series
\begin{eqnarray}
\label{equation_generic_double_series}
h({x}_1, x_{2}) = e^{-{x_{2}}} \sum_{n_{1}=0}^{\infty} \sum_{{n_{2}}=0}^\infty \frac{(N)_{n_{1}}}{(\NR + n_2)_{n_{1}}} \frac{{x}_1^{n_{1}}}{n_{1}!} \frac{ {x_{2}^{n_{2}}}}{{n_{2}}!} H_{n_{1}},
\end{eqnarray}
where $ H_{n_1} $ stands for $ M_{n_{1}} (s) $ from~(\ref{equation_mgf_gamma1_double_series_Gamma_mixture}), $ p_{n_{1}} (t) $ from~(\ref{equation_gamma1_pdf_double_series_Gamma_mixture}), $ P_{\text{o},n_{1}} $ from~(\ref{equation_Po_double_series_Gamma_mixture}), and $ C_{n_1} $ from~(\ref{equation_capacity_double_series_Gamma_mixture}).
Thus, the dependence of $ h({x}_1, x_{2}) $ on $ s $ for the m.g.f.~or $ t $ for the p.d.f.~is not explicitly shown in~(\ref{equation_generic_double_series}), for simplicity.

Numerical results not shown due to length limitations have revealed that increasing $ K $, $ \NR $, and $ \NT $ yield increasingly problematic numerical convergence for series~(\ref{equation_generic_double_series}). 
This is explained by: 1) the fact that~(\ref{equation_mgf_gamma1_double_series_Gamma_mixture}) has been obtained from~(\ref{equation_mgf_gamma1_uncond_main_result}) by replacing $  {\hFoo}\!\left(N; n_{2} + \NR;  \frac{ \Gamma_1 s}{1- \Gamma_1 s} {x}_1 \right) $ with its expansion around $ x_1 = 0 $ from~(\ref{equation_1F1_series}); 2) the fact that $ x_1 $ is increasing because of the following proportionality, proved in Appendix~\ref{section_relationship_x1_x2}:
\begin{eqnarray}
\label{equation_x1_detail}
x_1 \propto K \NR \NT.
\end{eqnarray}
%
Appendix~\ref{section_relationship_x1_x2} also shows that $ x_2 \propto K \NR \NT $. Then, the expressions for $ x_1 $ and $ x_2 $ deduced there in~(\ref{equation_x1_detail_again}) and~(\ref{equation_x2_detail_again}) can be used to show that their ratio $ c_1 = \frac{x_1}{x_2} $ is real-valued, positive, and independent of $ K $ and $ \NR $.
Finally, unshown numerical results have revealed that $ c_1 \propto 1/\NT $.
These considerations suggest substituting $ x_2 = z $ and $ x_1 = c_1 z $ in the generic series in~(\ref{equation_generic_double_series}), which yields the following result.

\begin{lemma}
\label{lemma_single_infinite_series}
For $ x_2 = z $ and $ x_1 = c_1 z $, series~(\ref{equation_generic_double_series}) can be recast as the single-infinite series
\begin{eqnarray}
\label{equation_generic_single_series}
h(z) = e^{-z} \sum_{n=0}^{\infty} \underbrace{\sum_{m=0}^{n} \binom{n}{m}  \frac{(N)_{m}}{(\NR + n - m)_{m}}H_{m} c_1^{m}}_{=G_n} \frac{z^n}{n!}.
\end{eqnarray}
Derivatives of $ h(z) $, required below for HGM, are given by
\begin{eqnarray}
\label{equation_generic_single_series_derivatives}
\partial_z^k h (z)= \sum_{l = 0}^{k} \binom{k}{l} (-1)^{k - l} e^{-z} \sum_{n=l}^{\infty} {G}_{n} \frac{z^{n-l}}{(n-l)!}.
\end{eqnarray}
\end{lemma}

\begin{IEEEproof}
The proof of the first part is not shown, due to simplicity and length limitations.
The second part follows from~(\ref{equation_generic_single_series}) based on Leibniz's formula\cite[Eq.~(1.4.12), p.~5]{NIST_book_10}.
\end{IEEEproof}



Numerical results shown later reveal that the truncation of~(\ref{equation_generic_single_series}) still does not converge numerically for practically relevant values of $ K $, $ \NR $, and $ \NT $. 
Therefore, we shall endeavor to compute it by HGM, as done for Rician--Rayleigh fading in\cite{siriteanu_twc_14} to compute the {\sf SNR}  p.d.f.~series deduced from~(\ref{equation_mgf_gamma1_Rice_Ray_sum}). 
Recall that HGM evaluates a function at given values for its variables by numerically solving its differential equations starting from initial conditions, i.e., known values of the function and required derivatives, at another point\cite[Sec.~IV.B]{siriteanu_twc_14}.
Thus, HGM requires differential equations.
Note that, making the substitutions $ x_2 = z $ and $ x_1 = c_1 z $ and regarding $ c_1 $ as a constant factor, conveniently reduces the number of variables in generic series~(\ref{equation_generic_single_series}).
For example, when cast for the m.g.f., the series is only a function of $ s $ and $ z $.

Differential equations were derived by hand, with difficulty, for the {\sf ZF SNR} m.g.f.~and p.d.f.~in\cite[Eqs.~(32), (42)]{siriteanu_twc_14} for the Rician--Rayleigh fading case, based on the {\sf SNR} m.g.f.~expression shown here in~(\ref{equation_mgf_gamma1_Rice_Ray}) and the differential equation satisfied by $ {\hFoo} (N; \NR; \sigma) $, i.e.,\cite[Eq.~(27)]{siriteanu_twc_14}
\begin{eqnarray}
\label{equation_1F1_differential_equation}
\sigma \cdot \hFoob (N; \NR; \sigma) + (\NR - \sigma) \cdot \hFooa (N; \NR; \sigma) - N \cdot  \hFoo (N; \NR; \sigma) = 0.
\end{eqnarray}

For the full-Rician fading case with $ r = 1 $ studied herein, the new {\sf SNR} m.g.f.~expression in~(\ref{equation_mgf_gamma1_uncond_main_result}) comprises an extra sum compared to~(\ref{equation_mgf_gamma1_Rice_Ray}). On the other hand,~(\ref{equation_generic_single_series}) yields the following complicated  {\sf SNR} m.g.f.~expression:
\begin{eqnarray}
\label{equation_generic_single_series_MGF}
M_{\gamma_1} (s; z) \!\!\!\! & = & \!\!\!\! e^{-z} \sum_{n=0}^{\infty} {\sum_{m=0}^{n} \binom{n}{m}  \frac{(N)_{m} M_{m} (s) c_1^{m} }{(\NR + n - m)_{m}} } \frac{z^n}{n!}, \quad \\
M_{m} (s) \!\!\!\! & = & \!\!\!\! \sum_{{m_1} = 0}^{m} \binom{m}{{m_1}} (-1)^{m_1} \frac{1}{(1 - s \Gamma_1)^{N + m - {m_1}}}. \quad 
\end{eqnarray}
Because the by-hand derivation of differential equations w.r.t.~$ s $ and $ z $ satisfied by $ M_{\gamma_1} (s; z) $ described by~(\ref{equation_mgf_gamma1_uncond_main_result}) or~(\ref{equation_generic_single_series_MGF}) is not tractable, we shall apply instead the automated approach described below, based on the generic expression~(\ref{equation_generic_single_series}). 
The derivation of differential equations satisfied by $ p_{\gamma_1} (t; z)  $, $ P_{\text{o}}(z) $, and $ C(z) $ can be automated as well, based on: 1) their generic expression~(\ref{equation_generic_single_series}); or 2) the Laplace-transform relationship between $ M_{\gamma_1} (s; z) $ and $ p_{\gamma_1} (t; z)  $, and the integral relationships of $ p_{\gamma_1} (t; z)  $ with $ P_{\text{o}}(z) $ and $ C(z) $. We shall employ the latter approach because it is more general.

\section{Computer-Algebra-Aided Derivation of Differential Equations for HGM}

\label{section_differential_equations_all}

\subsection{Holonomic Functions, Annihilator, {Gr{\"o}bner Basis}, and Creative Telescoping}
A function is \textit{holonomic} w.r.t. a set of continuous variables if it satisfies for
each of them a linear differential equation with polynomial
coefficients. A
function is holonomic w.r.t. to a set of discrete variables if the associated
generating function is holonomic in the previous sense\cite[Sec.~IV.C]{siriteanu_twc_14}\cite[p.~17]{koutschan_thesis_09}.   
For example, $
{\hFoo} (N; \NR; \sigma) $ is holonomic w.r.t.~$\sigma$ because it
satisfies differential equation\footnote{Note that $
{\hFoo} (N; \NR; \sigma) $ is also holonomic w.r.t.~$N$ and~$\NR$.}~(\ref{equation_1F1_differential_equation}).
In other words, $ {\hFoo} (N; \NR; \sigma)
$ is \textit{annihilated} by the differential operator $ \sigma \partial_\sigma^2 + (\NR - \sigma)
\partial_\sigma - N $. 
The (infinite) set of all operators that annihilate a
given holonomic function is called its \textit{annihilator}\cite[p.~18]{koutschan_thesis_09}. 

Holonomic functions are closed under addition, multiplication, certain
substitutions, and taking sums and integrals\cite{siriteanu_twc_14},\cite{koutschan_thesis_09}. 
Consequently, functions $ M_{\gamma_1} (s; z) $,
$ p_{\gamma_1} (t; z) $, $ P_{\text{o}}(z) $, and $ C(z) $, cast as
in~(\ref{equation_generic_single_series}), are holonomic. 
The fact that the
closure properties for holonomic functions can be executed algorithmically provides a
systematic way of deriving the differential equations required for HGM, by
starting with the annihilating operators of the comprised ``elementary''
holonomic functions in~(\ref{equation_generic_single_series}).
A key ingredient for algorithmically executing closure properties is the algebraic concept of
\textit{Gr{\"o}bner basis}, which provides a canonical and finite representation of an annihilator and helps decide whether an operator is in an annihilator.
For details on \textit{Gr{\"o}bner bases} theory, computation, and applications see\cite{buchberger_thesis_65},\cite{koutschan_thesis_09},\cite{geng_jsac_14},\cite{fozunbal_tit_04},\cite{lin_tcs_08} and references therein. 


While many holonomic closure properties require, basically, only linear algebra, computing the
annihilator for a sum or integral of a holonomic function is a more
involved task. For example, one can employ the \textit{creative telescoping} technique: given an integral $ F(x) = \int_a^b f(x, y)\,\mathrm{d}y $, creative telescoping algorithmically finds in the annihilator of $f(x,y)$
a differential operator of the form $P(x,\partial_x)+\partial_y\cdot
Q(x,y,\partial_x,\partial_y)$. 
Then, using the
fundamental theorem of calculus\cite[p.~6]{NIST_book_10} and differentiating under the integral sign reveals\footnote{Under ``natural boundary''
conditions\cite{koutschan_thesis_09}.} $P(x,\partial_x) $ as an annihilating operator for~$F(x)$\cite[p.~46]{koutschan_thesis_09}. 
Several creative telescoping algorithms are described in\cite[Ch.~3]{koutschan_thesis_09}. 

\subsection{The \texttt{HolonomicFunctions} Computer-Algebra Package}

This freely-available computer-algebra package, written earlier in \texttt{Mathematica}
by one of the authors, is described, with numerous
examples, in\cite{koutschan_tr_10b}.  Its commands implement: 1) the
computation of {Gr{\"o}bner bases} in operator algebras, 2) closure properties
for holonomic functions, and 3) creative telescoping algorithms
from\cite[Ch.~3]{koutschan_thesis_09}.
Thus, it enables automated deduction of
differential equations for holonomic functions (e.g., our m.g.f.~infinite
series), their Laplace transform (e.g., our p.d.f.), and their integrals
(e.g., our outage probability and ergodic capacity). Conveniently, its
symbolic-computation ability\footnote{Inherited from \texttt{Mathematica}.}
allows for parameters (e.g., $ \NR $, $ N $, $ \Gamma_1 $, $ \tau $, $
c_1 $).

\subsection{Computer-Algebra-Aided Derivation}

The \texttt{Mathematica} file with \texttt{HolonomicFunctions} commands that produce the output discussed and employed below can be downloaded from\cite{koutchan_Mathnb_LOS}. 
Therein, for example, {Gr{\"o}bner basis} computation with the command \texttt{Annihilator} yields annihilating operators for expression $  e^{-z} \frac{z^n}{n!} $ from~(\ref{equation_generic_single_series}).
Further, the command \texttt{CreativeTelescoping} yields annihilating operators for $ G_n $ based on its definition as the inner sum in~(\ref{equation_generic_single_series}), and for $P_{\text{o}}(z)$ based on the integral in~(\ref{equation_Po_definition}).

Note that the particular functions that enter the differential equations shown below --- i.e., $ M_{\gamma_1} (s; z) $, $ \partial_{s} M_{\gamma_1} (s; z) $, $ \partial_{z} M_{\gamma_1} (s; z) $; $ p_{\gamma_1} (t; z)$, $ \partial_{t} p_{\gamma_1} (t; z) $, $ \partial_{z} p_{\gamma_1} (t; z) $, $ \partial_z^2  p_{\gamma_1} (t; z) $; $ \partial_{z}^{k} P_{\text{o}}(z) $, $ k = 0:4 $; $ \partial_{z}^{k} C(z) $, $ k = 0:6 $ --- arise automatically from~(\ref{equation_generic_single_series}) by {Gr{\"o}bner basis} computation and creative telescoping, and are revealed with the command \texttt{UnderTheStaircase} in\cite{koutchan_Mathnb_LOS}.

The steps and outcomes of the procedure implemented by the code in\cite{koutchan_Mathnb_LOS} are as follows:
\begin{enumerate}
\item Derive {\sf SNR}  m.g.f.~differential equations w.r.t.~$ s $ and $ z $, based on~(\ref{equation_generic_single_series}). Then,\cite{koutchan_Mathnb_LOS} reveals that the function vector
\begin{eqnarray}
\label{equation_vector_mgf_derivatives}
\um (s; z) =\begin{matrix} (M_{\gamma_1} (s; z) & \partial_{s} M_{\gamma_1} (s; z) & \partial_{z} M_{\gamma_1} (s; z) )^{\sf T} \end{matrix} \nonumber
\end{eqnarray}satisfies the systems of differential equations w.r.t.~$s$ and $ z $
\begin{eqnarray}
\label{equation_diffeq_mgf_wrt_s}
\label{equation_diffeq_mgf_wrt_z}
\partial_s \um (s; z) = \uTheta_{s} \um (s; z), \;
\partial_{z} \um (s; z) = \uTheta_{z} \um (s; z),
\end{eqnarray}
with the $ 3 \times 3 $ matrices $  \uTheta_{s} $ and $  \uTheta_{z} $ shown only in\cite{koutchan_Mathnb_LOS}, due to space limitations.

\item Using results from Step 1, derive p.d.f.~differential equations w.r.t.~$ t $ and $ z $, based on the inverse-Laplace transform. Then,\cite{koutchan_Mathnb_LOS} reveals that the function vector
\begin{eqnarray}
\label{equation_vector_pdf_derivatives}
& \!\!\!\! \up (t; z) =\begin{matrix} (p_{\gamma_1} (t; z) & \!\!\! \partial_{t} p_{\gamma_1} (t; z) & \!\!\!  \partial_{z} p_{\gamma_1} (t; z)& \!\!\!  \partial_z^2  p_{\gamma_1} (t; z))^{\sf T} \end{matrix} \nonumber
\end{eqnarray}
satisfies the systems of differential equations w.r.t.~$t$ and $ z $
\begin{eqnarray}
\label{equation_diffeq_pdf_wrt_t}
\label{equation_diffeq_pdf_wrt_z}
\partial_t \up (t; z) = \uXi_{t} \up (t; z), \quad
\partial_{z} \up (t; z) = \uXi_{z} \up (t; z),
\end{eqnarray}
with the $ 4 \times 4 $ matrices $  \uXi_{t} $ and $  \uXi_{z} $ shown in\cite{koutchan_Mathnb_LOS}.

\item Using results from Step 2, derive differential equations w.r.t.~$ z $ for $ P_{\text{o}}(z) $ and $ C(z) $, based on their integral relationships from~(\ref{equation_Po_definition}) and~(\ref{equation_ergodic_capacity_definition}) with $ p_{\gamma_1} (t; z) $. Then,\cite{koutchan_Mathnb_LOS} reveals that the function vectors $ \up_{\text{o}} (z) \doteq  5 \times 1 $ with $ [\up_{\text{o}} (z)]_k = \partial_{z}^{k} P_{\text{o}}(z) $, $ k = 0:4 $, and $ \uc (z) \doteq  7 \times 1 $ with $ [\uc (z) ]_k = \partial_{z}^{k} C(z)  $, $ k = 0:6 $, 
satisfy the systems of differential equations
\begin{eqnarray}
\label{equation_diffeq_Po_wrt_z}
\partial_{z} \up_{\text{o}} (z) = \uPhi_{z} \up_{\text{o}} (z), \quad
\label{equation_diffeq_ergodic_capacity_wrt_z}
\partial_{z} \uc (z) = \uPsi_{z} \uc (z),
\end{eqnarray}
where $ \uPhi_{z} \doteq 5 \times 5 $ and  $  \uPsi_{z} \doteq  7 \times 7 $ are companion matrices\cite[p.~109]{gentle_book_07} shown in\cite{koutchan_Mathnb_LOS}.

The above systems of differential equations enable the HGM-based computation of the {\sf SNR}  p.d.f., outage probability, and ergodic capacity, as shown below.


\end{enumerate}

\section{Numerical Results}
\label{section_numerical_results}

\subsection{Description of Parameter Settings and Approaches}
\label{section_parameters_settings}

For the channel-matrix mean in~(\ref{equation_Hd_outer_product}), unit-norm vector $ \ua $ and vector $ \ub $ with the norm in~(\ref{equation_b_norm_squared}) are constructed, according to\cite[Eq.~(7.29), p.~299]{tse_book_05}, from array response vectors\footnote{See\cite[Fig. 7.3b, p. 296, Eq. (7.20), p. 297]{tse_book_05} for geometry and derivation details.}, as
\begin{eqnarray}
\label{equation_a_ARV_normalized}
\ua \!\!\!\! & = & \!\!\!\!  \frac{1}{\sqrt{\NR}} 
\begin{matrix} (1 & e^{-j  \pi \cos(\theta_{\text{R}})} & \dots & e^{-j \pi (\NR - 1) \cos(\theta_{\text{R}})})^{\sf T} \end{matrix}, \quad \\
\label{equation_b_ARV}
\ub \!\!\!\!  & = & \!\!\!\!  \frac{1}{\sqrt{\NT}} 
\begin{matrix} (1 & e^{-j  \pi  \cos(\theta_{\text{T}})} & \dots & e^{-j \pi (\NT - 1) \cos(\theta_{\text{T}})})^{\sf T} \end{matrix}   \sqrt{\frac{K}{K+1} \NR \NT},
\end{eqnarray}
assuming uniform linear antenna arrays with interelement spacing of half of the carrier wavelength. 
Above, $ \theta_{\text{R}} $ and $ \theta_{\text{T}} $ are, respectively, the angles of arrival and departure of the LoS component w.r.t.~the antenna broadside directions. 
Unless stated otherwise, we assume $ \theta_{\text{R}} = 30^\circ $ and $ \theta_{\text{T}} $ equal to the central angle, $ \theta_{\text{c}} $, of the transmit-side Laplacian power azimuth spectrum\cite[Eq.~(4.2)]{siriteanu_thesis_06}. 
Correlation matrix $ \uRT $ is computed from the AS and $ \theta_{\text{c}} $ with\cite[Eqs.~(4-3)--(4-5)]{siriteanu_thesis_06}.

Section~\ref{section_results_outage} below shows results for the Stream-1 outage probability for $ \tau = 8.2 $~dB, which corresponds to a symbol error probability of $ 10^{-2} $ for QPSK modulation. Thus, the constellation size is $ M = 4 $, and we show $ P_{\text{o}} $ vs.~$ \Gamma_{\text{b}} =  \Gamma_{\text{s}}/\log_2 M = \Gamma_{\text{s}}/2 $.
On the other hand, Section~\ref{section_results_capacity} shows results for the sum rate, i.e., the sum of the ergodic capacities of all streams, in bits per channel use (bpcu), vs.~AS, $ K $, and $  \theta_{\text{T}} $. Also shown are simulation results for maximum-likelihood detection ({\sf ML}).

Unless stated otherwise, presented results have been obtained by running \texttt{MATLAB R2012a}, in its native fixed precision, on a computer with a 3.4-GHz, 64-bit, quad-core\footnote{Nevertheless, we have run single instances of \texttt{MATLAB} when measuring the computation time (with \texttt{tic}, \texttt{toc}.)} processor and 8 GB of memory.
For the simulation results (in figure legends: \texttt{Sim.}) we have employed, when feasible, $ N_{\text{s}} = 10^6 $ samples of $ \un $ and $ \uH $ for~(\ref{equation_received_signal}), to produce reliable results for $ P_{\text{o}} $ as low as $ 10^{-5} $.
Then, series results (in legends: \texttt{Series}) have been produced by truncating~(\ref{equation_generic_single_series}) as in\cite[Section~V.F]{siriteanu_twc_13}, i.e., new terms have been added until: 1) their relative change falls below $10^{-10} $, or 2) $ n \le n_{\text{max}} = 150$, as additional terms in~(\ref{equation_generic_single_series}) lead to numerical divergence because the arising large numbers are represented with poor precision.
Numerical divergence is indicated in legends with \texttt{Series$^*$}.
Outage probability results for full-Rayleigh fading (in legend: \texttt{Rayleigh,Exp.}) have been obtained with expression $ P_{\text{o}}  = \frac{\gamma \left( N,  \tau/\Gamma_1 \right)}{(N - 1)!} $, obtained from~(\ref{equation_Po_double_series_Gamma_mixture}) based on Remark~\ref{remark_special_case_x2}.
Finally, HGM results (in legends: \texttt{HGM}) have been produced by solving --- with the \texttt{MATLAB} \texttt{ode45} function with tolerance levels of $ 10^{-10} $ --- the systems of differential equations in~(\ref{equation_diffeq_Po_wrt_z}).
Then, for the outage probability, the initial condition $ \up_{\text{o}} ( z_0 ) $ has been computed accurately with~(\ref{equation_generic_single_series}) and~(\ref{equation_generic_single_series_derivatives}) at $ z_0 = 0.05692 $, which arises from~(\ref{equation_x2_Gd2_norm}) for $ K = -25 $~dB, $ \NR = 6 $, $ \NT = 4 $, and $ \uRT = \mbI_{\NT} $. 
Finally, sum rate results have been obtained by adding the ergodic capacities of the $ \NT $  streams.

Results are shown for $ K $ and AS values relevant to their lognormal distributions for WINNER II scenarios A1 (indoors office) and C2  (urban macrocell), under LoS propagation\cite[Table~5.5]{winner_d_1_1_2_v_1_2}: 1) averages of these distributions, i.e., for $ K = 7 $~dB, and for $ \text{AS} = 51^\circ $ and $ 11^\circ $, which yield low and  high antenna element correlation, i.e., $ | [\uRT]_{1, 2} | = 0.12 $ and $ 0.83 $, respectively; 2) values within the range of most likely values\cite[Table~1]{siriteanu_tvt_11}, or 3) random samples\footnote{Then, even computing $ \uRT $ with\cite[Eqs.~(4-3)--(4-5)]{siriteanu_thesis_06} is time consuming; nevertheless, the employed $ 2,100 $ samples of AS and $ K $ have yielded smooth outage probability plots.}.


\subsection{Outage Probability Results}
\label{section_results_outage}

\subsubsection{Description of Results for $K$ and AS Relevant to Scenario A1, and for Small $ \NR $ and $ \NT $}

\begin{figure}[t]
\begin{center}
\includegraphics[width=5in]
{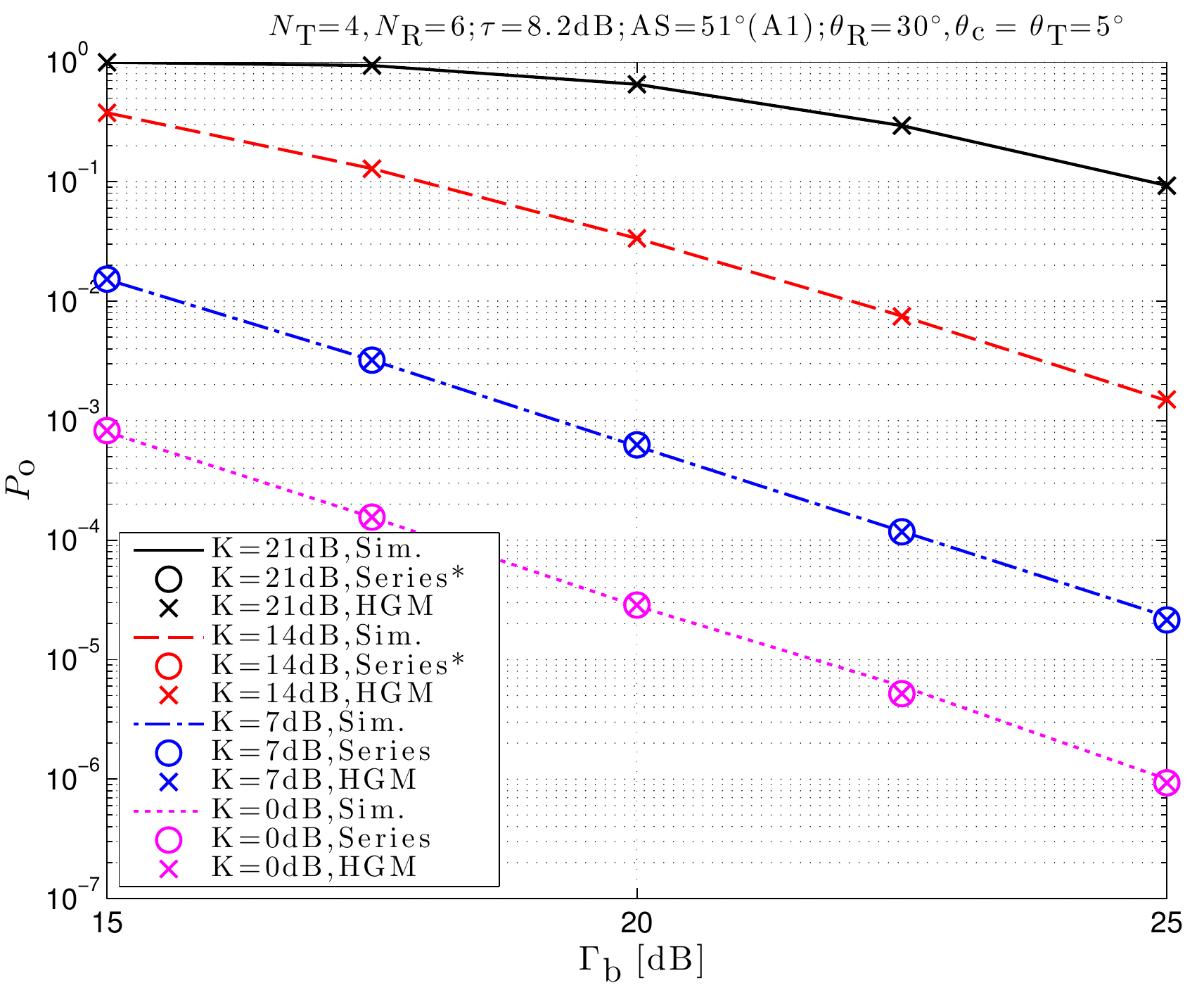}
\caption{Stream-1 outage probability for $ \NR = 6 $, $ \NT = 4 $, $ \text{AS} = 51^\circ $ (i.e., scenario A1 mean), and various values of $ K $, including $ K = 7 $~dB (i.e., scenario A1 mean). Series results for $ K=14, 21$~dB do not appear because of numerical divergence.}
\label{figure_ZF_LOS_NR_6_NT_4_Po_vs_SNR_sev_K_A1_AS_51}
\end{center}
\end{figure}

Fig.~\ref{figure_ZF_LOS_NR_6_NT_4_Po_vs_SNR_sev_K_A1_AS_51} shows results for $ \text{AS} = 51^\circ $ and $K$ set to values from $ 0 $~dB to the upper limit of the range expected with $ 0.99 $ probability for scenario A1\cite[Table~1]{siriteanu_tvt_11}. Note that the \texttt{MATLAB} series truncation diverges for $ K = 14 $~dB and $ 21 $~dB\footnote{Our series truncation in \texttt{Mathematica}, with its arbitrary precision, converged also for $ K = 14 $~dB, but required one hour vs.~a few seconds for HGM; series truncation in \texttt{Mathematica} was not tried for $ K = 21 $~dB.}, whereas HGM and simulation results agree at all $ K $. 
Thus, HGM enables us to investigate the performance degradation likely to occur in practice with increasing $ K $ for MIMO {\sf ZF} under full-Rician fading with $ r=1 $.


%

\begin{figure}[t]
\begin{center}
\includegraphics[width=5in]
{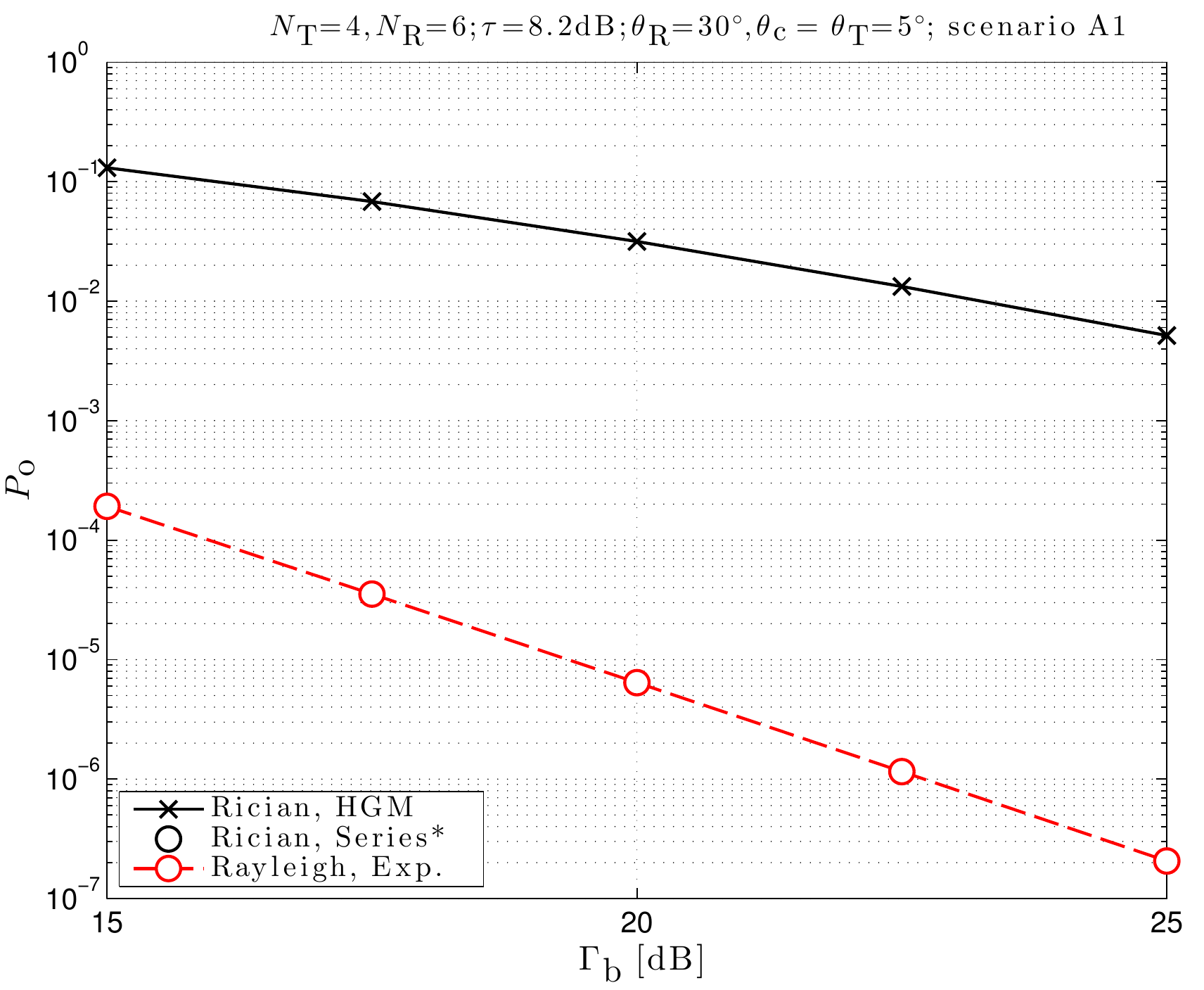}
\caption{Stream-1 outage probability for $ \NR = 6 $, $ \NT = 4 $, averaged also over the WINNER II lognormal distributions of $ K  $ and $ \text{AS}$ for scenario A1. Results corresponding to \texttt{Rician,Series}  do not appear because of numerical divergence.}
\label{figure_ZF_LOS_NR_6_NT_4_Po_vs_SNR_rand_K_AS_A1}
\end{center}
\end{figure}

Fig.~\ref{figure_ZF_LOS_NR_6_NT_4_Po_vs_SNR_rand_K_AS_A1} shows results from averaging also over AS and $ K $ from their WINNER II lognormal distributions for scenario A1. 
First, simulation has not been attempted due to the long required time. (The computation time is explored in more detail below.) Series truncation does not yield useful results because of numerical divergence for the larger $ K $ values.
Only HGM has yielded relatively expeditiously a smooth plot whose unshown continuation at sufficiently large $ \Gamma_{\text{b}} $ has revealed the expected diversity order\footnote{The expected diversity order is also noticeable from the plots for $ K = 0 $~dB and $ 7 $~dB in Fig.~\ref{figure_ZF_LOS_NR_6_NT_4_Po_vs_SNR_sev_K_A1_AS_51}.} of $ N = 3 $\cite[Eq.~(46)]{siriteanu_twc_13}.

Figs.~\ref{figure_ZF_LOS_NR_6_NT_4_Po_vs_SNR_sev_K_A1_AS_51} and~\ref{figure_ZF_LOS_NR_6_NT_4_Po_vs_SNR_rand_K_AS_A1} depict the same $ \Gamma_{\text{b}} $ range in order to reveal that: 1) setting AS and $ K $ to their averages can substantially overestimate performance vs.~averaging over AS and $ K $ --- compare the blue dash-dotted plot in Fig.~\ref{figure_ZF_LOS_NR_6_NT_4_Po_vs_SNR_sev_K_A1_AS_51} with the solid black plot in Fig.~\ref{figure_ZF_LOS_NR_6_NT_4_Po_vs_SNR_rand_K_AS_A1}; 2) making the assumption of full-Rayleigh fading instead of full-Rician fading leads to unrealistic performance expectations --- compare the plots in Fig.~\ref{figure_ZF_LOS_NR_6_NT_4_Po_vs_SNR_rand_K_AS_A1}.

\subsubsection{Description of Results for $K$, AS Relevant to Scenarios A1, C2, and for Increasing $ \NR $, $ \NT $}

\begin{center}
\begin{table*}[t]
{\small
\hfill{}
\caption{Results for $ K = 7 $~{dB}, $ \text{AS} = 51^\circ$ (i.e., scenario A1) and $ \text{AS} = 11^\circ$ (C2), and $ (\NR,\NT) = N_a \times (6, 4)$.}
\renewcommand{\arraystretch}{0.5}
\label{table_Massive_MIMO_results}
\begin{tabular}{|c|c|c|c|c|c|c|c|}
  \hline
   AS & $ N_a $ & $ \Gamma_{\text{b}}$ (dB)  & $ P_{\text{o}} = [a \!\times\! 10^{-2}, b \!\times\! 10^{-5} ] $ & Series & Sim. ($ N_{\text{s}} = 10^6 $) & HGM \\
   \hline \hline
   $ 51^\circ$ (A1) & 1 & [15, 25] & $ a = 1.53 , b = 2.15 $ & 1.3 s \cmark & 31 s & 20 s \cmark \\
   \hline
      $ 51^\circ$ (A1) & 2 & [11, 17] & $ a = 1.74, b = 4.26 $ & 1.3 s \xmark & 53 s& 20 s \cmark \\
   \hline
      $ 51^\circ$ (A1) & 5 & [6, 9] & $ a = 1.39, b = 6.39 $ & 1.3 s \xmark & 520 s& 20 s \cmark \\
   \hline
      $ 51^\circ$ (A1) & 10 & [2, 4.5] & $ a = 2.35, b = 2.45 $ & 1.3 s \xmark & 2,300 s& 20 s \cmark \\
   \hline
      $ 51^\circ$ (A1) & 15 & [0, 2] & $ a = 1.98, b = 1.61 $ & 1.3 s \xmark & 8,800 s& 20 s \cmark \\
   \hline
      $ 51^\circ$ (A1) & 100 & [-9.2, -8.5] & $ a = 2.72, b = 2.57 $ & 1.3 s \xmark &  $ estimated: 1.9 \times 10^6$ s~\xmark & 20 s \cmark \\
      \hline \hline
   $ 11^\circ$ (C2) & 1 & [23, 32] & $ a = 1.12 , b = 3.01 $ & 1.3 s \cmark & 31 s & 20 s \cmark \\
      \hline 
   $ 11^\circ$ (C2) & 2 & [18.5, 24.5] & $ a = 1.43 , b = 3.36 $ & 1.3 s \xmark & 54 s & 20 s \cmark \\
      \hline 
   $ 11^\circ$ (C2) & 10 & [5, 7.5] & $ a = 2.12 , b = 2.09 $ & 1.3 s \xmark & 2,400 s & 20 s \cmark \\
   \hline 
\end{tabular}}
\end{table*}
\end{center}

Table~\ref{table_Massive_MIMO_results} summarizes results of several numerical experiments for $ K $ and AS set to their averages for scenarios A1 and C2, and for the pair $ (\NR,\NT) $ set to $ N_a \times (6, 4)$, with $ N_a $ shown in the second column\footnote{Note that $ \NR $ does not necessarily have to be much larger than $ \NT $ even in massive MIMO\cite{bjornsson_cm_15}.}.
The $ \Gamma_{\text{b}} $ ranges shown in the third column yield $ P_{\text{o}} $ in the order of $ 10^{-2}$ -- $10^{-5}$, as shown in the fourth column.
The remaining three columns show the actual or estimated computation time (in seconds), per $ \Gamma_{\text{b}} $ value.
The marks \cmark~and \xmark~in the `Series' column denote, respectively, successful and unsuccessful (i.e., numerical divergence) series computation\footnote{For $ (\NR = 6, \NT =4) $ numerical convergence is achieved with $ n = 134 $, whereas the other $ (\NR, \NT ) $ pairs yield $ n =  n_{\text{max}}= 150 $. Consequently, \texttt{MATLAB} reports about the same computation time ($ \approx 1.3 $~s) for all cases.}.
Further, mark \xmark~in the `Sim.' column indicates infeasible simulation duration.
Finally, mark \cmark~in the `HGM' column indicates successful HGM-based computation.
This table demonstrates that, unlike series truncation and simulation, HGM enables reliable, accurate, and expeditious {\sf ZF} assessments for realistic $ K $ and even large MIMO. 

\begin{figure}[t]
\begin{center}
\includegraphics[width=5in]
{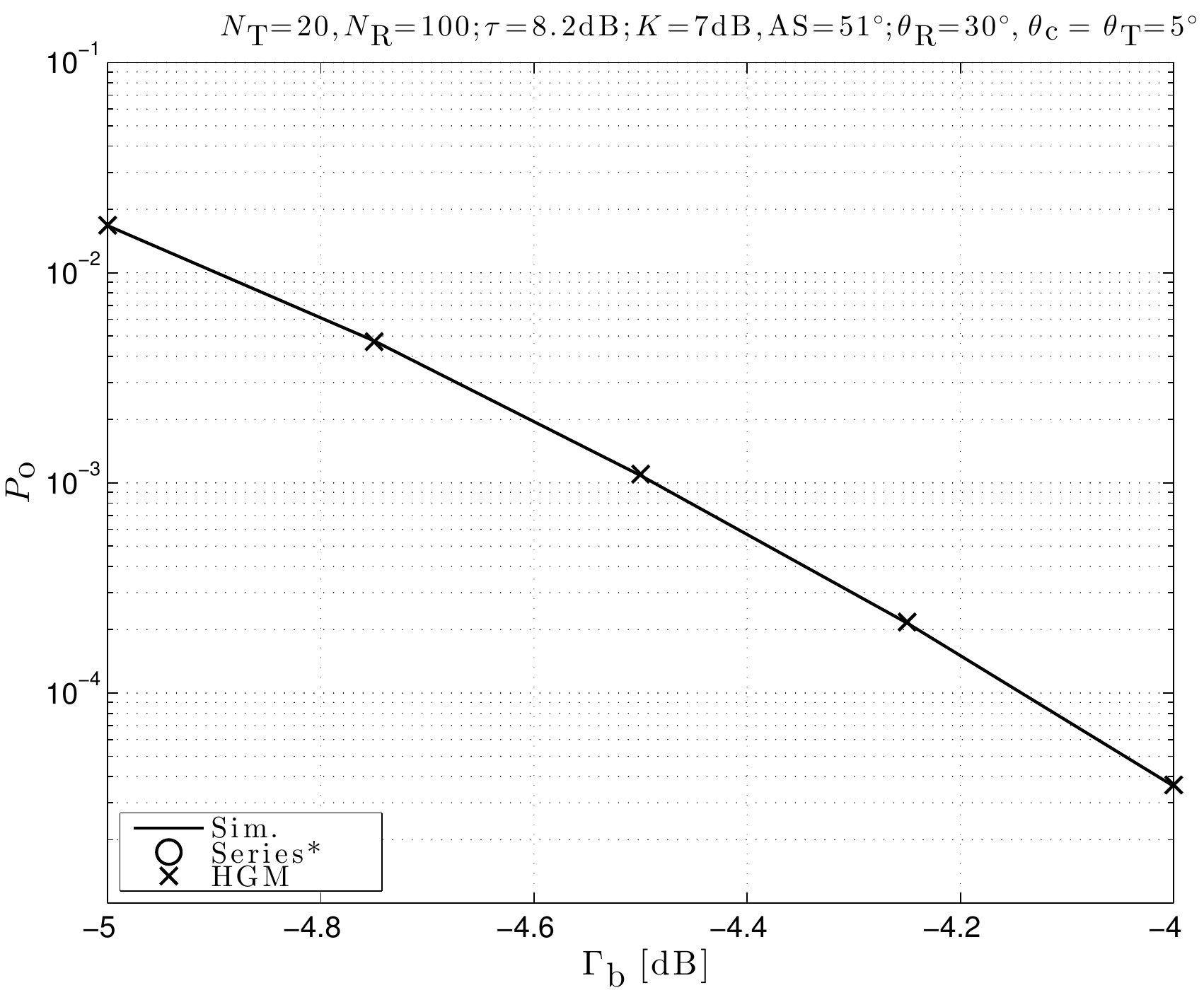}
\caption{Stream-1 outage probability for $ \NR = 100 $, $ \NT = 20 $, for  $ K = 7 $~dB and $ \text{AS} = 51^\circ $ (i.e., averages for scenario A1). Results corresponding to \texttt{Series}  do not appear because of numerical divergence.}
\label{figure_ZF_LOS_NR_100_NT_20_Po_vs_SNR_K_7_A1_AS_51}
\end{center}
\end{figure}

Fig.~\ref{figure_ZF_LOS_NR_100_NT_20_Po_vs_SNR_K_7_A1_AS_51} characterizes {\sf ZF} performance for $ K = 7 $~dB and $ \text{AS} = 51^\circ $, and for the large-MIMO setting with $ \NR = 100 $ and $ \NT = 20 $. 
On the one hand, series truncation does not produce useful results; on the other hand, HGM results agree with the simulation results, and we have found HGM over 30 times faster\footnote{When large $ \NT $ yields infeasibly-long simulation, HGM results can be validated by checking the diversity order revealed by its $ P_{\text{o}} $-vs.-$ \Gamma_{\text{b}} $ plot. E.g., for $ \NR = 104 $ and $\NT= 100 $, we have found its slope magnitude to be near the expected $ N = 5 $.}.

\subsection{Ergodic Capacity Results}
\label{section_results_capacity}

The ZF ergodic capacity has been computed, for each stream, for $ \NR = 6 $, $ \NT = 4 $, $ \theta_{\text{c}} = 5^\circ $, and $ \Gamma_{\text{s}} = 10 $~dB by: 1) HGM based on~(\ref{equation_diffeq_ergodic_capacity_wrt_z}) with $ \uPsi_{z} $ shown in\cite{koutchan_Mathnb_LOS}, 2) simulation (also for {\sf ML}), and 3) the infinite series in~(\ref{equation_generic_single_series}). 
Results from the series do not appear in the figures because, as for the outage probability, its truncation diverges for realistic values of $ K $.

\begin{figure}
\begin{center}
\includegraphics[width=5in]
{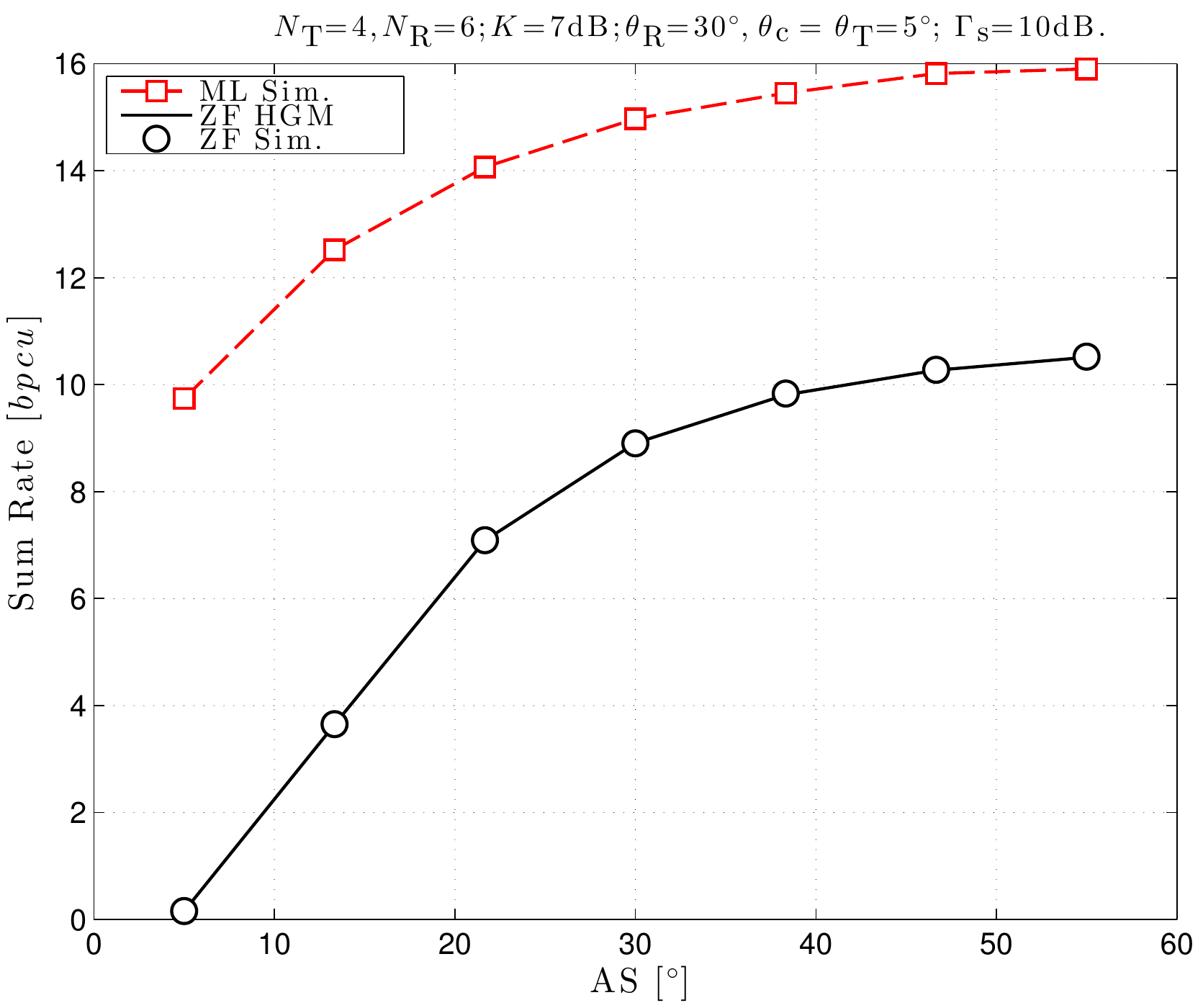}
\caption{{\sf ZF} sum rate from HGM and simulation vs.~AS, for $ \NR = 6 $, $ \NR = 4 $, $ K = 7 $~dB; also, {\sf ML}  sum rate from simulation.}
\label{figure_ZF_NR_6_NT_4_C_vs_AS_A1_K_7_Sum}
\end{center}
\end{figure}

\begin{figure}
\begin{center}
\includegraphics[width=5in]
{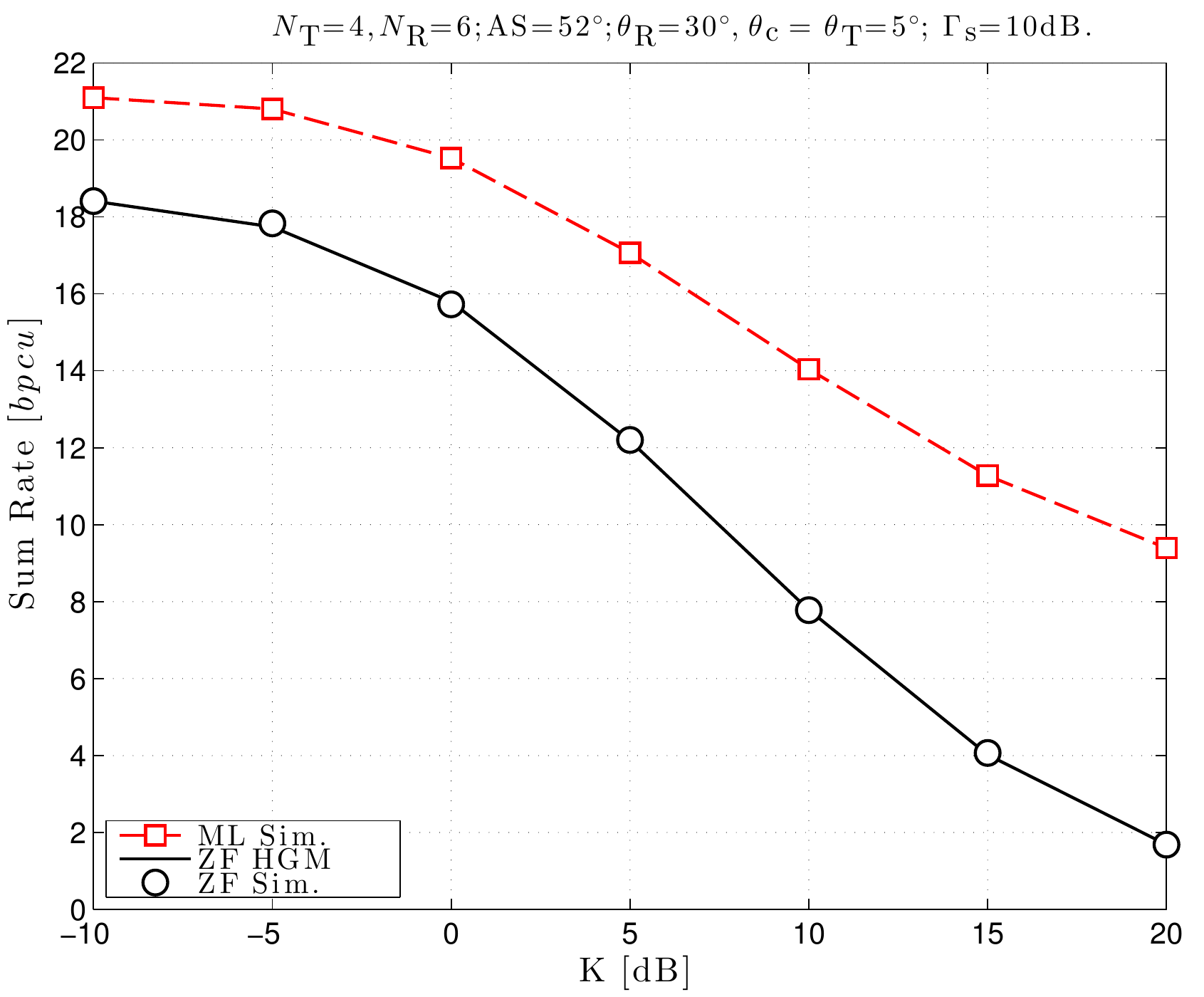}
\caption{{\sf ZF} sum rate from HGM and simulation vs.~$ K $, for $ \NR = 6 $, $ \NR = 4 $, $ \text{AS} = 52^\circ $; also, {\sf ML}  sum rate from simulation.}
\label{figure_ZF_C_vs_K_NT_4_NR_6_Rice_AS_52_Sum}
\end{center}
\end{figure}

Fig.~\ref{figure_ZF_NR_6_NT_4_C_vs_AS_A1_K_7_Sum} demonstrates that increasing AS (decreasing antenna correlation) yields increasing {\sf ZF} sum rate and decreasing {\sf ML}--{\sf ZF} rate gap. 
On the other hand, Fig.~\ref{figure_ZF_C_vs_K_NT_4_NR_6_Rice_AS_52_Sum} reveals that increasing $ K $ yields decreasing {\sf ZF} sum rate and increasing {\sf ML}--{\sf ZF} rate gap, for large AS (e.g., $ 51^\circ $). 
However, other (unshown) results indicate that the {\sf ML}--{\sf ZF} gap is decreasing for small AS (e.g., $ 7^\circ $) and is constant for medium AS (e.g., $ 12^\circ $).

\begin{figure}
\begin{center}
\includegraphics[width=5in]
{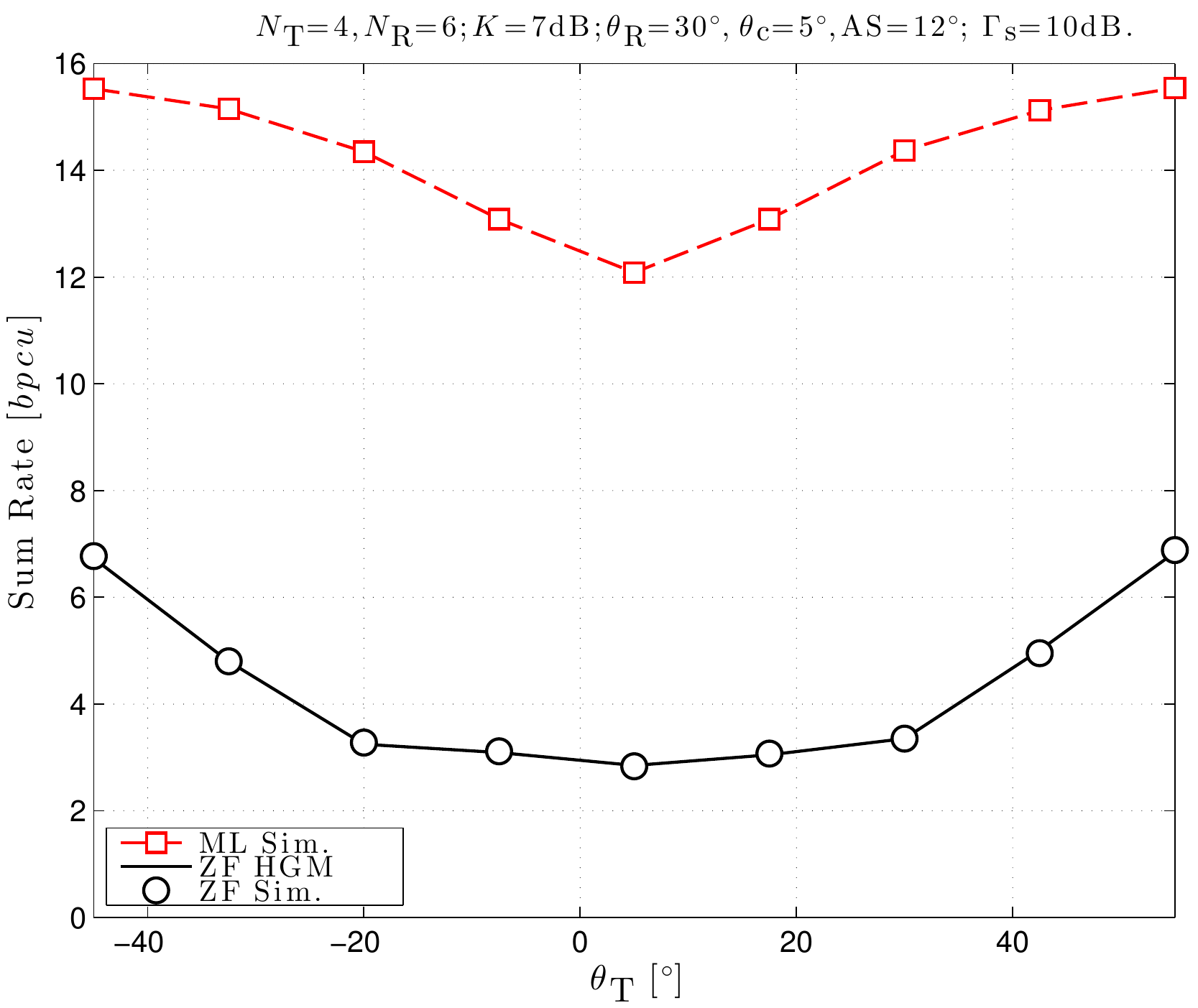}
\caption{Sum rate vs.~$ \theta_{\text{T}} $ for $ \theta_{\text{c}} = 5^\circ $, when $ \NR = 6 $, $ \NR = 4 $, $ K = 7 $~dB, $ \text{AS} = 12^\circ $; also, {\sf ML}  sum rate from simulation.}
\label{figure_ZF_C_vs_Tht_NT_4_NR_6_C2_K_7_AS_12_Sum}
\end{center}
\end{figure}

Finally, Fig.~\ref{figure_ZF_C_vs_Tht_NT_4_NR_6_C2_K_7_AS_12_Sum} reveals, for $ \text{AS} = 12^\circ $ and $ \theta_{\text{c}} = 5^\circ $, a substantial sum rate decrease with decreasing $ | \theta_{\text{T}} - \theta_{\text{c}} | $.
Based on Remark~\ref{remark_SNR_statistics}, because condition $ \theta_{\text{T}} = \theta_{\text{c}} $ yields worst performance, it must also minimize $ \| \uhda - \uHdb \ur_{2,1} \| $.
For larger AS, other (unshown) results have revealed more moderate rate gain with increasing  $ | \theta_{\text{T}} - \theta_{\text{c}} | $.
For very large AS (e.g., $ 51^\circ $), the sum rate remains unchanged with increasing $ | \theta_{\text{T}} - \theta_{\text{c}} | $, because large AS yields $ \ur_{2,1} \approx \mzero $, i.e., $ \| \uhda - \uHdb \ur_{2,1} \| \approx \| \uhda \| $, which is independent of $ | \theta_{\text{T}} - \theta_{\text{c}} | $.
Unshown numerical results from the approximating gamma distribution from Remark~\ref{remark_Rayleigh_approx} have revealed it inaccurate especially for small $ \NR $, $ \NT $, and $ K $. 
On the other hand, we have found that accuracy improves with smaller $ | \theta_{\text{T}} - \theta_{\text{c}} | $, which corroborates Remark~\ref{remark_SNR_statistics}.

\section{Summary, Conclusions, and Future Work}

This paper has provided an exact performance analysis and evaluation of MIMO spatial multiplexing with ZF, under transmit-correlated full-Rician fading with LoS component of rank $ r = 1 $.
First, we expressed as infinite series the {\sf SNR}  m.g.f.~and p.d.f., as well as performance measures, e.g., the outage probability and ergodic capacity.
However, their numerical convergence has been revealed inherently more problematic with increasing $ K $, $ \NR $, and $ \NT $.
Therefore, we have applied computer algebra to the derived infinite series and deduced satisfied differential equations.
They have been used for HGM-based computation. 
Thus, we have expeditiously produced accurate results for the range of realistic values of $ K $ and even for large $ \NR $ and $ \NT $.
Consequently, we have been able to assess the substantial performance degradation incurred with increasing $ K $ for {\sf ZF} when $ r=1 $. 
Furthermore, HGM has helped reveal that the performance averaged over WINNER II AS and $ K $ distributions can be much worse than that for average AS and $ K $.
Finally, we have been able to evaluate the performance for antenna numbers relevant to large MIMO reliably and much more expeditiously than by simulation.

Based on our experience studying MIMO for Rician fading for {\sf ZF} in this paper and for {\sf MRC}\footnote{Here, MRC refers to the MIMO technique of transmitting and receiving over the dominant channel mode, as discussed in\cite{kang_jsac_03}.} in our ongoing work, we expect that performance measure expressions for larger $ r $ and other transceiver techniques shall entail multiple infinite series in factors proportional to $ K $, $ \NR $ and $ \NT $, which shall diverge numerically for realistic values of these parameters. 
Alternate computation with the HGM shall require differential equations.
Because their by-hand derivation from the infinite series shall be intractable, computer algebra shall be indispensable.

\appendix

\subsection{Proof of Lemma~\ref{lemma_U211_distribution}}
\label{lemma_2}



Based on~(\ref{equation_G2_definition}) and~(\ref{equation_Gd2_elements}), we can regard $ [\uGb]_{\bullet,1} \doteq \NR \times 1 $,  as a vector of independent complex-valued Gaussians with variance of $ 1/2 $ for the real and imaginary parts, and means 
\begin{eqnarray}
\mathbb{E}\{[\uGb]_{1,1}\}={\sqrt{x_{2}}}, \quad \mathbb{E}\{[\uGb]_{i, 1}\}=0,  \; i = 2 : \NR,
\end{eqnarray}
which yield
\begin{eqnarray}
\label{equation_G211_distribution}
\frac{|[\uGb]_{1,1}|^2}{1/2} & \sim & \chi^2_2 \left(\frac{{x_{2}}}{1/2}\right), \\
\label{equation_G2i1_distribution}
\frac{|[\uGb]_{2,1}|^2}{1/2} + \dots + \frac{| [\uGb]_{\NR,1} |^2}{1/2} & \sim &  \chi^2_{2 (\NR - 1)}.
\end{eqnarray}
Now, because $ \uT_2 $ in~(\ref{equation_G2_QR_decomposition}) is upper triangular, we can write the first column of $  \uGb = \uUb \uT_2 $ as $ [\uGb]_{\bullet,1} = [\uUb]_{\bullet,1} [\uT_2]_{11} $.
If we set
\begin{eqnarray}
\label{equation_U2_col1}
[\uT_2]_{1,1} = \Vert [\uGb]_{\bullet,1} \Vert, \quad [\uUb]_{\bullet,1} = \frac{[\uGb]_{\bullet,1}}{\Vert [\uGb]_{\bullet,1} \Vert}, 
\end{eqnarray}
then
\begin{eqnarray}
|[\uUb]_{1,1}|^2 = \frac{|[\uGb]_{1,1}|^2}{ |[\uGb]_{1,1}|^2 + |[\uGb]_{2,1}|^2 + \dots + | [\uGb]_{\NR,1} |^2 }.
\end{eqnarray}
Finally, using~(\ref{equation_G211_distribution}),~(\ref{equation_G2i1_distribution}), and the independence of $ [\uGb]_{i, 1} $, $ i = 1 : \NR $, one can show
that\cite{chattamvelli_as_95}
\begin{eqnarray}
\label{equation_beta2_distribution}
|[\uUb]_{1,1}|^2 & \sim & {\sf B}(1, \NR - 1, 2 {x_{2}}), \nonumber \\
\label{equation_beta3_distribution}
\beta_1  { \,{\buildrel {(\ref{equation_Q211_as_product})} \over =}\, } 1 - |[\uUb]_{1,1}|^2 & \sim & {\sf B}(\NR - 1, 1, 2 {x_{2}}). \nonumber
\end{eqnarray}

The p.d.f.~of $\beta_1 $ is then given by\cite{chattamvelli_as_95}  
\begin{equation}
\label{equation_beta_3_noncentral}
f_{\beta_1}(v) = \sum_{{n_{2}}=0}^\infty \frac{e^{-x_2} x_2^{n_2}}{n_2!} \underbrace{ \left( \frac{ v^{(\NR - 1)-1}(1 - v)^{({n_{2}}+1)-1}}{\int_{0}^{1}  t^{(\NR - 1)-1} (1-t)^{({n_{2}}+1)-1} \,\mathrm{d} t} \right)}_{ =f_{\beta_3}(v; \NR-1, {n_{2}}+1)},
\end{equation}
where $f_{\beta_3}(v; \NR-1, {n_{2}}+1)$ is the p.d.f.~of a variable $ \beta_3 \sim {\sf B}(\NR-1, {n_{2}}+1)$.
Then, the $n_{1}$th moment of $ \beta_1 $ is
\begin{eqnarray}
\label{equation_U211_moments}
 \mathbb{E} \{ \beta_1^{n_1} \} =
 \sum_{{n_{2}}=0}^\infty \frac{e^{-x_2} x_2^{n_2}}{n_2!} \mathbb{E}\{ \beta_3^{n_{1}} \} = \sum_{{n_{2}}=0}^\infty  \frac{ e^{-{x_{2}}} {x_{2}^{n_{2}}}}{{n_{2}}!} \frac{(\NR -1)_{n_{1}}}{(n_{2} + \NR)_{n_{1}}}. 
\end{eqnarray}

\subsection{Proof of Lemma~\ref{lemma_rho_distribution}}
\label{lemma_1}

First, let us consider the  $ \NR \times \NR $  matrix $ \widehat{\uG}_2 =
\begin{matrix} (\uGb & \uGbtilde) \end{matrix}  \sim {\cal{CN}}_{\NR,\NR} \left( \widehat{\uG}_{\text{d},2}, \mb \mbI_{\NR} \otimes \mbI_{\NR} \right)$ obtained by joining the $ \NR \times \NI $ matrix $ \uGb \sim {\cal{CN}}_{\NR,\NI} \left( \uGdb, \mb \mbI_{\NR} \otimes \mbI_{\NI} \right) $ from~(\ref{equation_G2_definition}) --- whose  sole nonzero-mean column is $ [\uGb]_{\bullet,1} $ --- 
with the $ \NR \times N $ matrix $ \uGbtilde \sim {\cal{CN}}_{\NR, N} \left( \mzero, \mb \mbI_{\NR} \otimes \mbI_{N} \right) $.
Then, paralleling~(\ref{equation_G2_QR_decomposition}), let us consider its QR decomposition, i.e.,
\begin{eqnarray}
\label{equation_QR_decom_G2_hat}
\widehat{\uG}_2 =
\begin{matrix} (\uGb & \uGbtilde) \end{matrix} = \widehat{\uU}_2  \widehat{\uT}_2,
\end{eqnarray}
with $ \widehat{\uU}_2 \doteq \NR \times \NR $ unitary, i.e., $ \widehat{\uU}_2^{\sf H} \widehat{\uU}_2 = \widehat{\uU}_2 \widehat{\uU}_2^{\sf H} = \mbI_{\NR} $, and $ \widehat{\uT}_2 \doteq \NR \times \NR $ upper triangular with positive diagonal elements.
By partitioning in~(\ref{equation_QR_decom_G2_hat}) and using~(\ref{equation_G2_QR_decomposition}), we can write
\begin{eqnarray} 
\label{equation_G2_hat_partitioning}
\widehat{\uG}_2 =
\begin{matrix} (\uGb & \uGbtilde) \end{matrix}  = \begin{matrix} (\uUb & \uUbtilde) \end{matrix}  \left(
  \begin{array}{cc}
    \uT_2 &  \widetilde{\uT}_{12} \\
     \mzero &  \widetilde{\uT}_{22} \\
  \end{array}
  \right) =\begin{matrix} (\uUb \uT_2 & \uUb \widetilde{\uT}_{12} + \uUbtilde  \widetilde{\uT}_{22}) \end{matrix},
\end{eqnarray}
where 
$ \uUbtilde \doteq \NR \times N $ satisfies $ \uUbtilde^{\sf H} \uUbtilde = \mbI_{N} $, 
$ \widetilde{\uT}_{12} \doteq \NI \times N $,
and
$ \widetilde{\uT}_{22} \doteq N \times N $ is upper triangular with positive diagonal elements.

Hereafter, let us assume that $ [\uGb]_{\bullet,1} $ is given, i.e., $[\uUb]_{\bullet,1}$ set as in~(\ref{equation_U2_col1}) is given.
Then, the distribution of 
\begin{eqnarray}
\widehat{\uG}_2 &
{ \,{\buildrel {(\ref{equation_QR_decom_G2_hat})} \over = }\, }  &
\widehat{\uU}_2  \widehat{\uT}_2 = ( \begin{matrix} \uUb & \uUbtilde \end{matrix}) \widehat{\uT}_2 \nonumber \\ & = & 
 ( \begin{matrix}[\uUb]_{\bullet,1} &[\uUb]_{\bullet,2} & \dots & [\uUb]_{\bullet,\NI} & \uUbtilde \end{matrix}) \widehat{\uT}_2 \nonumber
\end{eqnarray}
is invariant to unitary transformations of the columns $ [\uUb]_{\bullet,i} $, $ \forall i = 2 : \NI $ and the columns of $ \uUbtilde $. Thus, we may rewrite
\begin{eqnarray}
\label{equation_partitioning_U2_hat}
\widehat{\uU}_2 = \begin{matrix}  (\uUb & \uUbtilde) \end{matrix} = \begin{matrix}  \big([\uUb]_{\bullet,1} & \uU^0 \uP \big) \end{matrix}, \quad 
\end{eqnarray}
where $ \uU^0 \doteq \NR \times (\NR - 1) $ comprises fixed orthonormal vectors selected to form a basis with $ [\uUb]_{\bullet,1} $, and 
$\uP \doteq (\NR-1) \! \! \times \! \! (\NR-1) $ is unitary, Haar-distributed\cite[Sec.~III.E]{siriteanu_twc_13}, not dependent on $[\uUb]_{\bullet,1}$.  
Using the first row of $\uU^0 $
to define
\begin{eqnarray}
\label{equation_q_definition_Gaussian}
\uqsimple^{\sf T} = [\uU^0]_{1,\bullet} \cdot \uP \doteq 1 \times (\NR - 1),
\end{eqnarray}
the first row of $ \widehat{\uU}_2 $ from~(\ref{equation_partitioning_U2_hat}) can be written as
\begin{eqnarray}
\label{equation_U2_first_row}
[\widehat{\uU}_2]_{1, \bullet} = \begin{matrix} ([\uUb]_{1,1} & \uqsimple^{\sf T}) \end{matrix}.
\end{eqnarray}
Then, based on $ \widehat{\uU}_2 \widehat{\uU}_2^{\sf H} = \mbI_{\NR} $ and~(\ref{equation_U2_first_row}), we can write
\begin{eqnarray}
\label{equation_qnorm_one_minusUnorm}
1 =  \| [\widehat{\uU}_2]_{1, \bullet} \|^2 = |[\uUb]_{1,1}|^2 + \| \uqsimple \|^2 \Rightarrow \| \uqsimple \|^2 = 1 - |[\uUb]_{1,1}|^2.
\end{eqnarray}
From~(\ref{equation_q_definition_Gaussian}) and~(\ref{equation_qnorm_one_minusUnorm}) we deduce that the vector
\begin{eqnarray}
\frac{\uqsimple}{ \| \uqsimple \|} = \frac{\uqsimple}{\sqrt{1 - |[\uUb]_{1,1}|^2}}
\end{eqnarray}
is uniformly distributed on the unit sphere $ \mathbb{S}^{\NR-2}$. 

Finally, because we can write
\begin{eqnarray}
 \label{equation_U2_first_row_detail}
  [\widehat{\uU}_2]_{1, \bullet} & { \,{\buildrel {(\ref{equation_partitioning_U2_hat})} \over = }\, }  & \begin{matrix}  ([\uUb]_{1,1} & [\uUb]_{1,2}  & \dots & [\uUb]_{1,\NI}  & [\uUbtilde]_{1,\bullet}) \end{matrix} \nonumber \\ 
  & { \,{\buildrel {(\ref{equation_U2_first_row})} \over = }\, }  & \begin{matrix} ([\uUb]_{1,1} & q_1 & \dots & q_{\NI -1} & q_{\NI} & \dots & q_{\NR -1}) \end{matrix},\nonumber
\end{eqnarray}
we have that $[\uUb]_{1,2}, \dots, [\uUb]_{1,\NI}$ are the first $\NI-1$ elements of $ \uqsimple $. 
Thus, we can write, by also using~(\ref{equation_qnorm_one_minusUnorm}),
\begin{eqnarray}
\label{equation_beta1_distr}
{\beta_4} & = & \frac{ |[\uUb]_{1,2}|^2 + \dots + |[\uUb]_{1,\NI}|^2}{1 - |[\uUb]_{1,1}|^2} \nonumber \\
& = & \frac{|q_1|^2 + \dots + |q_{\NI-1}|^2}{(|q_1|^2 + \dots + |q_{\NI-1}|^2) + (|q_{\NI}|^2  + \dots + |q_{\NR-1}|^2 )}. \nonumber
\end{eqnarray}
Recalling that $\frac{\uqsimple}{ \| \uqsimple \|}$ is uniformly distributed, we can deduce that, conditioned on $ [\uGb]_{\bullet,1} $, i.e., on $ [\uUb]_{\bullet,1} $, random variables $ {\beta_4} $ and $ {\beta_2} \,{\buildrel {(\ref{equation_Q211_as_product})} \over =}\, 1 - \beta_4 $ have the following distributions\cite{chattamvelli_as_95}:
\begin{eqnarray}
{\beta_4} & \sim & {\sf B}(\NI-1, \NR - \NI ) = {\sf B}(\NT-2, N), \nonumber \\
\label{equation_rho_beta_proved}
{\beta_2} = 1 - \beta_4 &  \sim & {\sf B}(\NR - \NI, \NI-1) = {\sf B}(N, \NT-2). \quad \quad \nonumber
\end{eqnarray}
Because the distribution of $ \beta_2 $ does not depend on $ [\uUb]_{\bullet,1} $, we also deduce that $ {\beta_2} $ is independent of $ \beta_1  { \,{\buildrel {(\ref{equation_Q211_as_product})} \over =}\, } 1 - |[\uUb]_{1,1}|^2 $.

\subsection{Derivation of Expressions for $ x_1 $ and $ x_2 $}
\label{section_relationship_x1_x2}
From Remark~\ref{remark_b_norm}, the normalized vector $ \ub_{\text{n}} = \frac{\ub}{\|\ub\|} \doteq \NT \times 1 $
does not depend on $ K $.
Defining
\begin{eqnarray}
\label{equation_R_tilde}
\widetilde{\uR} & = &  \left(
  \begin{array}{cc}
    0 & \mzero \\
    \mzero & \RTbbinva \\
  \end{array}
  \right) \doteq \NT \times \NT, \\
\label{equation_r21_tilde}
\urtilde_{2,1} & = & \begin{matrix} (1 & - \ur_{2,1}^{\sf T})^{\sf T} \end{matrix} \doteq \NT \times 1,
\end{eqnarray}
we can write $ \mu_1 $ from~(\ref{equation_umu_simple_form}) and $ \ubtilde^{\sf H} \RTbbinva \ubtilde $ from~(\ref{equation_x2_Gd2_norm}) as:
\begin{eqnarray}
\label{equation_mu1_inner_product}
&& \mu_1 = b_1^* - \ubtilde^{\sf H} \ur_{2,1} = \ub^{\sf H} \urtilde_{2,1} = \| \ub \| \, \ub_{\text{n}}^{\sf H} \urtilde_{2,1}, \\
\label{equation_btilde_Hermitian_form}
&& \ubtilde^{\sf H} \RTbbinva \ubtilde = \ub^{\sf H} \widetilde{\uR} \ub = \| \ub \|^2  \, \ub_{\text{n}}^{\sf H} \widetilde{\uR} \ub_{\text{n}}.
\end{eqnarray}

Finally, from~(\ref{equation_b_norm_squared}) we have that $ \| \ub \|^2 = K \NR \NT/(K+1) $. From~(\ref{equation_RTK_Hr_RT}) we have that $ \RTinvaa \propto (K + 1) $ and $ \RTbbinva \propto (K + 1) $, i.e., $ \widetilde{\uR} \propto (K + 1) $, whereas $ \ur_{2,1} $ defined in~(\ref{equation_r21}), i.e., $ \urtilde_{2,1}  $ defined in~(\ref{equation_r21_tilde}), does not depend on $ K $.
These yield:
\begin{eqnarray}
\label{equation_x1_detail_again}
x_1 & { \,{\buildrel {(\ref{equation_x1_scalar})} \over =}\, } & \RTinvaa | \mu_1 |^2 \propto K \NR \NT, \quad \\
\label{equation_x2_detail_again}
x_2 & { \,{\buildrel {(\ref{equation_x2_Gd2_norm})} \over =}\, } &  \ubtilde^{\sf H} \RTbbinva \ubtilde \propto K \NR \NT.
\end{eqnarray}


\footnotesize
\bibliographystyle{IEEEtran}
\bibliography{IEEEabrv,books,journals,conferences,theses,miscellaneous}

\end{document}